\documentclass[twocolumn,english,aps,superscriptaddress,prl]{revtex4}
\usepackage{color}
\usepackage{amsmath}
\usepackage{amssymb}
\usepackage{graphicx}
\usepackage{dsfont}
\usepackage{ulem}

\newcommand{\w}{\omega}

\newcommand{\ii}{\imath}   

\newcommand{\HK}{\mathcal{H}_{\rm Kit}}
\newcommand{\HKK}{\mathcal{H}_{\rm KK}}
\newcommand{\HM}{\mathcal{H}_{\rm u}}

\newcommand{\HKo}{\mathcal{H}_{\rm Kon}}
\newcommand{\Hbath}{\mathcal{H}_{\rm bath}}
\newcommand{\Himp}{\mathcal{H}_{\rm imp}}
\newcommand{\Hhyb}{\mathcal{H}_{\rm hyb}}

\newcommand{\Simp}{S_{\rm imp}}
\newcommand{\cloc}{\chi_{\rm loc}}

\newcommand{\Cloc}{C_{\rm loc}}
\newcommand{\Ztwo}{\mathbb{Z}_2}
\newcommand{\Zthree}{\mathbb{Z}_3}

\begin{document}

\title{
Kondo impurities in the Kitaev spin liquid: \\
Numerical renormalization group solution and gauge-flux-driven screening
}

\author{Matthias Vojta}
\affiliation{Institut f\"ur Theoretische Physik, Technische Universit\"at Dresden, 01062 Dresden, Germany}
\author{Andrew K. Mitchell}
\affiliation{Institute for Theoretical Physics, Utrecht University, 3584 CE Utrecht, The Netherlands}
\author{Fabian Zschocke}
\affiliation{Institut f\"ur Theoretische Physik, Technische Universit\"at Dresden, 01062 Dresden, Germany}


\begin{abstract}
Kitaev's honeycomb-lattice compass model describes a spin liquid with emergent fractionalized excitations.
Here we study the physics of isolated magnetic impurities coupled to the Kitaev spin-liquid host.
We reformulate this Kondo-type problem in terms of a many-state quantum impurity coupled
to a multichannel bath of Majorana fermions and present the numerically exact
solution using Wilson's numerical renormalization group technique.
Quantum phase transitions occur as a function of Kondo coupling and locally applied field. At zero field, the impurity moment is partially screened only when it binds an emergent gauge flux, and otherwise becomes free at low temperatures.
We show how Majorana degrees of freedom determine the fixed-point properties, make contact with Kondo screening in pseudogap Fermi systems, and discuss effects away from the dilute limit.
\end{abstract}

\date{\today}


\maketitle

Spin liquids constitute a fascinating class of states of local-moment magnets, characterized by the absence of symmetry-breaking order at low temperatures \cite{balents10}. Very often these states are topologically non-trivial, displaying an emergent gauge structure and fractionalized excitations.
Given that the exotic properties of spin liquids are difficult to detect directly in local observables, much additional information can be obtained by studying the distinctive response to local perturbations. In particular, impurities or defects can act as \textit{in-situ} probes.
This general concept of characterizing non-trivial magnetic states via their response to isolated impurities has been successfully applied in the past, with prominent examples being vacancy-induced moments in confined spin-gap magnets \cite{vacmom}, universal fractional moments near quantum critical points \cite{sbv99}, the fractionalization of orphan spins in strongly frustrated magnets \cite{orphan}, and the pinning of emergent magnetic monopoles in spin ice \cite{mono_pin}.

In this paper we present a detailed study of the physics of a magnetic (Kondo) impurity coupled to the gapless spin-liquid phase of Kitaev's honeycomb-lattice compass model \cite{kitaev06}. This Kitaev model is a rare example of an exactly solvable model for a fractionalized spin liquid in two dimensions. Its solution can be cast into itinerant Majorana fermions coupled to a static $\Ztwo$ gauge field. These properties enable an exact reformulation of the Kondo problem in terms of a complex quantum impurity coupled to non-interacting fermions, suitable for treatment using Wilson's numerical renormalization group (NRG) \cite{nrg,nrg_rev}. NRG is a nonperturbative method that yields essentially exact numerical results down to lowest temperatures for any coupling strength.

\begin{figure}[t]
  \centering
    \includegraphics[width=8.4cm]{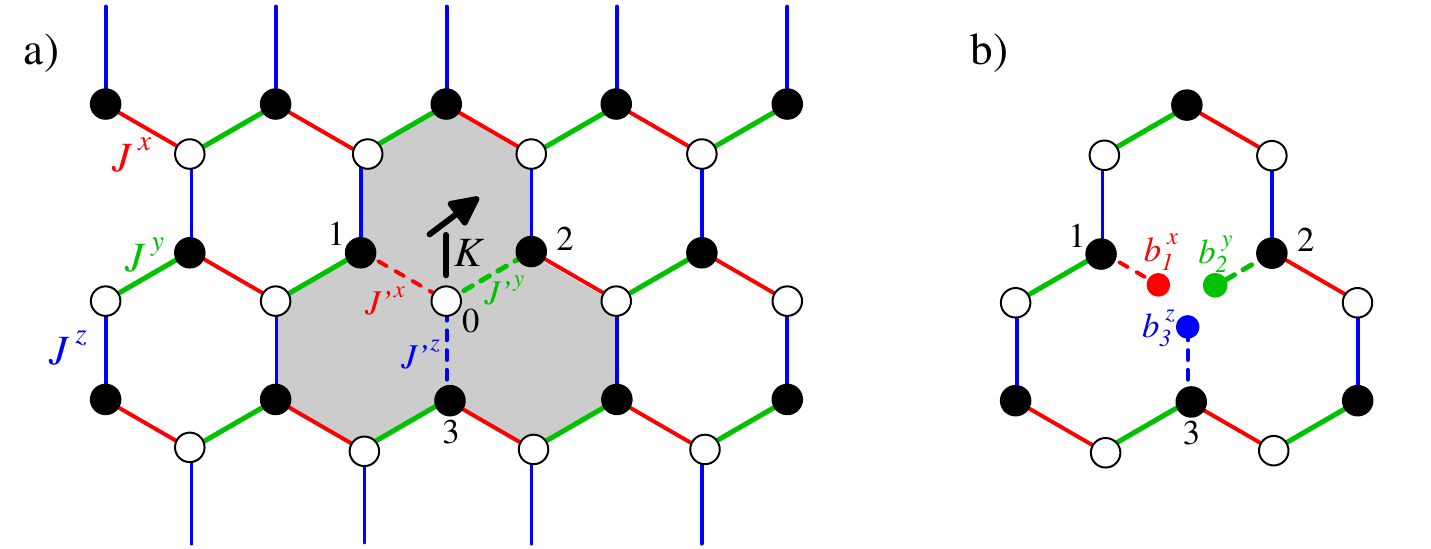}
\caption{
(a) Setup for the Kitaev Kondo problem: an extra spin (black arrow) and the bulk spin at site $0$ are coupled by the Kondo coupling $K\equiv K^x=K^y=K^z$. The bulk exchange couplings are denoted by $J^{x,y,z}$, and we allow for different couplings $J'^{x,y,z}$ next to the impurity. The large shaded plaquette will be dubbed ``impurity plaquette''.
(b) Illustration of dangling gauge Majorana modes relevant for the vacancy fixed points ($K\to\pm\infty$).
}
\label{fig:model}
\label{fig:imp}
\end{figure}

\begin{figure}[t]
\begin{center}
\includegraphics[width=8.6cm]{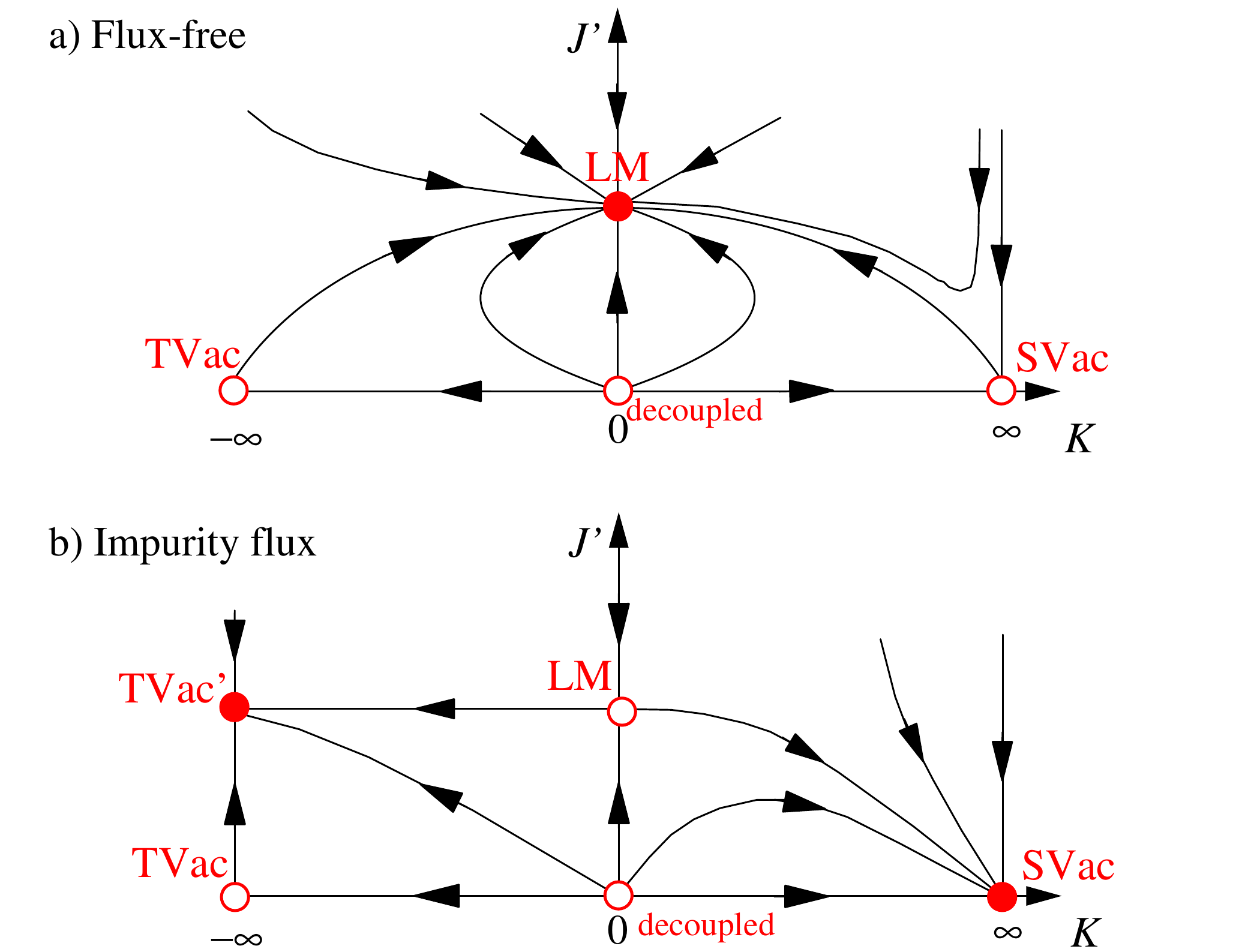}
\caption{\label{fig:flow1}
Schematic RG flow for the isotropic Kitaev Kondo model in a plane spanned by the Kondo coupling $K$ and the coupling $J'$ between site 0 and sites 1,2,3.
(a) Flux-free sector of the Hilbert space.
(b) Sector with a $\Ztwo$ flux through the impurity plaquette, $W_I=-1$.
Full (open) dots denote stable (unstable) fixed points.
}
\end{center}
\end{figure}

Our main results for an isolated Kondo impurity in the Kitaev spin liquid, Fig.~\ref{fig:model}, can be summarized as follows:
As function of the Kondo coupling $K$, the model displays a single first-order quantum phase transition (QPT) at $K=K_c>0$: There is partial screening for large antiferromagnetic couplings, $K_c<K<\infty$, whereas the impurity spin is unscreened otherwise, $-\infty<K<K_c$. The transition is accompanied by the binding of a gauge flux to the impurity. The renormalization-group (RG) flow in the individual flux sectors is non-trivial, see Fig.~\ref{fig:flow1}. Importantly, there is no screening -- and no QPT -- in the flux-free sector of the Hilbert space due to an emergent particle--hole symmetry. A magnetic field applied to the impurity can drive multiple transitions; it also induces flux binding for ferromagnetic $K$.
We are able to characterize all fixed points in terms of their magnetic response and residual entropy, in part arising from localized Majorana zero modes, and we provide analytical expressions for the relevant crossover scales.
Our results connect to those obtained for isolated vacancies \cite{willans10,willans11} and encompass the case of substitutional spin-$1$ impurities.

Our work goes far beyond previous approximate treatments of Kondo impurities in spin liquids  \cite{cassa,FFV06,kim08,dhochak10,dhochak15,ribeiro12}. For the Kitaev Kondo model, the full solution with NRG reveals a far richer range of physics, controlled by nonperturbative effects related to the pseudogap Kondo problem and not captured within weak-coupling RG schemes.
Our analysis corrects aspects of earlier work \cite{dhochak10,dhochak15} on the same model as detailed in Section IV of Ref.~\onlinecite{suppl}.


\textit{Model.}
The Kitaev model \cite{kitaev06} describes spins 1/2 at sites $i$ of a honeycomb lattice, with sublattices A and B. The Ising-like nearest-neighbor interactions $J^\alpha$, $\alpha =x,y,z$, are tied to the real-space bond direction, reflecting strong spin-orbit coupling.
The bulk Hamiltonian reads
\begin{equation}
\label{hk}
\HK =
-J^x \sum_{\langle ij\rangle_x} \hat{\sigma}_i^x \hat{\sigma}_j^x
-J^y \sum_{\langle ij\rangle_y} \hat{\sigma}_i^y \hat{\sigma}_j^y
-J^z \sum_{\langle ij\rangle_z} \hat{\sigma}_i^z \hat{\sigma}_j^z
\end{equation}
where $\hat{\sigma}_j^{\alpha}$ are Pauli matrices, and $\langle ij \rangle_\alpha$ denotes an $\alpha$ bond as in Fig. \ref{fig:model}.
We focus on the isotropic case $J\equiv J^x=J^y=J^z$.

We consider a Kondo problem with a spin-1/2 impurity, $\vec{S}$, coupled to the Kitaev spin on site 0 of sublattice A. The full Hamiltonian is $\HKK = \HK + \HKo$ with
\begin{equation}
\HKo =  \sum_\alpha K^\alpha \hat{S}^\alpha \hat{\sigma}_0^\alpha + \sum_\alpha h^\alpha \hat{S}^\alpha
\end{equation}
where $\vec{K}$ is the Kondo coupling and $\vec{h}$ a local field applied to the impurity \cite{sunityfoot}. For the purpose of analysis we will allow the exchange couplings connecting site $0$ to its neighbors $1,2,3$ to take values $J'^\alpha \neq J^\alpha$, see Fig.~\ref{fig:imp}.

\textit{NRG formulation.}
Application of NRG to the Kitaev Kondo problem is made possible because the bulk Kitaev model has an infinite number of conserved $\Ztwo$ fluxes: For every elementary plaquette with spins $1,\ldots,6$, the operator $\hat{W}_p = \hat{\sigma}_1^x \hat{\sigma}_2^y \hat{\sigma}_3^z \hat{\sigma}_4^x \hat{\sigma}_5^y \hat{\sigma}_6^z$ is conserved, with eigenvalues $W_p=\pm1$. Consequently, the Hilbert space decomposes into flux sectors, defined by the set of $\{W_p\}$.
Using the representation $\hat{\sigma}_i^\alpha=\ii \hat{b}_i^\alpha \hat{c}_i$ of each bulk spin in terms of four Majorana fermions, $\hat{c}_i$ and $\hat{b}_i^\alpha$ \cite{kitaev06}, the operators $\hat{u}_{ij} = \ii\hat{b}_i^{\alpha_{ij}}\hat{b}_j^{\alpha_{ij}}$, defined on each lattice bond, are separately conserved. Their eigenvalues $u_{ij}=\pm 1$ relate to the plaquette fluxes via $W_p = u_{21}u_{23}u_{43}u_{45}u_{65}u_{61}$.
For a given set $\{u_{ij}\}$ the original bulk Hamiltonian \eqref{hk} reduces to a tight-binding model for the $c$ (``matter'') Majorana fermions,
\begin{eqnarray}
\label{hm}
\HM = \ii\sum_{\langle ij\rangle_\alpha}J^\alpha u_{ij}\hat{c}_i \hat{c}_j\,,
\end{eqnarray}
with hopping energies $J^\alpha u_{ij}$ encoding the coupling to the static $\Ztwo$ gauge field. The ground state of $\HM$ is located in the flux-free sector, with $u_{ij}=1$, where the spectrum can be found by Fourier transformation \cite{kitaev06,suppl}.

In the presence of the Kondo term, $K\neq0$, the fluxes in the three plaquettes adjacent to site 0 are no longer individually conserved. However, their product $W_I$ (the flux in the impurity plaquette, Fig.~\ref{fig:imp}), as well as all outer fluxes, remain conserved. This implies that the bulk system, with site 0 removed, forms a bath of free fermions with Hamiltonian $\Hbath$ in any given flux sector. This bath is coupled to a generalized ``impurity'' which now consists of the Kondo spin and the Kitaev spin at site 0, including also the surrounding flux degrees of freedom. The impurity Hamiltonian $\Himp$ acts in a Hilbert space of 16 states. The coupling between impurity and bath, $\Hhyb$, arises from the Kitaev exchanges $J'^\alpha$ between site 0 and sites $1,2,3$, such that the bath is characterized by a $3\times3$ matrix propagator. For isotropic couplings and flux configurations preserving the $\Zthree$ lattice rotation symmetry, the bath can be decomposed into angular-momentum modes, such that $\Hbath$ eventually consists of three channels of spinless fermions. The explicit forms of $\Hbath$, $\Himp$, and $\Hhyb$ are specified in the supplement \cite{suppl}.

With fluxes fixed, the Hamiltonian $\Hbath+\Himp+\Hhyb$ is equivalent to $\HKK$ and can be solved via NRG \cite{nrg,nrg_rev}. Iterative diagonalization of a semi-infinite-chain Hamiltonian yields the many-particle level flow as well as physical observables as function of temperature. Since separate NRG runs are performed in each flux sector, NRG thermodynamics are representative of the full Kitaev model at temperatures below the flux gap. Based on the results in Refs.~\onlinecite{kitaev06,willans10} we expect the ground state of $\HKK$ to have a flux-free bath, and $W_I$ either $+1$ or $-1$.


\begin{figure}[t]
\begin{center}
\includegraphics[width=8.7cm]{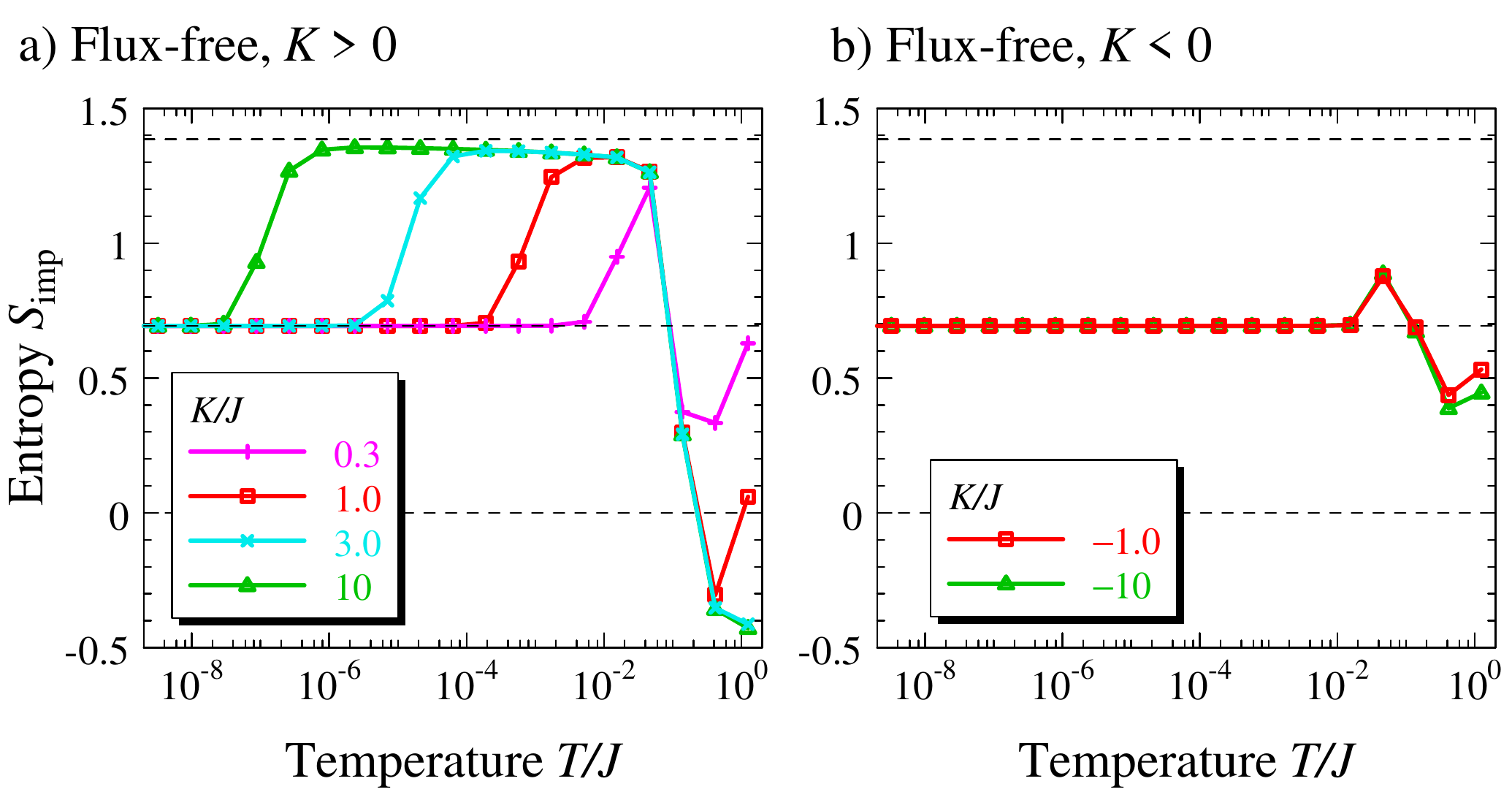}
\caption{\label{fig:s_noflux1}
NRG results for the impurity contribution to the total entropy, $\Simp(T)$, vs. $T/J$ in the no-flux case for $J'=J$ and various $K$. The horizontal dashed lines indicate $\Simp=0$, $\ln 2$, and $\ln 4$. The crossover behavior for $T/J>10^{-1}$ is a band-edge effect.
}
\end{center}
\end{figure}

\textit{Fixed points.}
We start by enumerating the trivial RG fixed points -- this is most efficiently done using the couplings $K$ and $J'$ as (renormalized) parameters.
$K=J'=0$ describes the fully decoupled situation, while $K=0$ and $J'=J$ is the local-moment fixed point (LM), where the unscreened Kondo spin is decoupled from the Kitaev bulk. Further, $K=+\infty$ causes singlet formation between the Kondo spin and the Kitaev spin 0, i.e., Kondo screening -- this induces a singlet vacancy (SVac) in the host (therefore $J'=0$). Similarly, $K=-\infty$ and $J'=0$ corresponds to a triplet vacancy (TVac).  We also find a fixed point at $K=-\infty$ but with finite $J'$, denoted TVac$'$. Note that all fixed points are separately defined in each flux sector.


\textit{NRG results: Flux-free case.}
We have extensively studied the Kitaev Kondo model using NRG. We start with results for the flux-free sector, i.e., $W_p=+1$ on all outer plaquettes and $W_I=+1$ on the impurity plaquette.

The impurity entropy $\Simp(T)$, obtained as the entropy difference between the systems with and without Kondo spin (with fluxes fixed), is shown in Fig.~\ref{fig:s_noflux1} for various values of $K/J$, keeping $J'=J$. We find that the impurity entropy reaches $\ln 2$ in the low-$T$ limit for all couplings $K$; the NRG level structure of this fixed point is identical to that at $K=0$, $J'=J$ \cite{suppl}. We conclude that in the flux-free case, there is a \textit{single stable phase} controlled by the LM fixed point, Fig.~\ref{fig:flow1}, without Kondo screening.

For $K\gtrsim J$, we find an intermediate crossover from $\Simp \approx \ln 4$ to $\ln 2$ upon lowering $T$. As explained below, the fixed point associated with $\ln 4$ entropy is identified as SVac, corresponding to a strong coupling Kondo-screened state. However, this fixed point is unstable: below a scale $T^\ast$, well fit by $T^\ast \propto J'^3 J^2 / K^4$ \cite{suppl}, the system evolves toward the LM fixed point and the impurity moment becomes free.

For ferromagnetic $K<0$ (Fig.~\ref{fig:s_noflux1}b) the LM fixed point is again stable. For $J'=J$, no intermediate RG flow is observed. Only for small $J'$ do we see incipient flow via TVac, with $\Simp \approx \ln 12$; the LM crossover scale in this case is extracted as $T^\ast \propto J'$ \cite{suppl}.



\begin{figure}[t]
\begin{center}
\includegraphics[width=8.7cm]{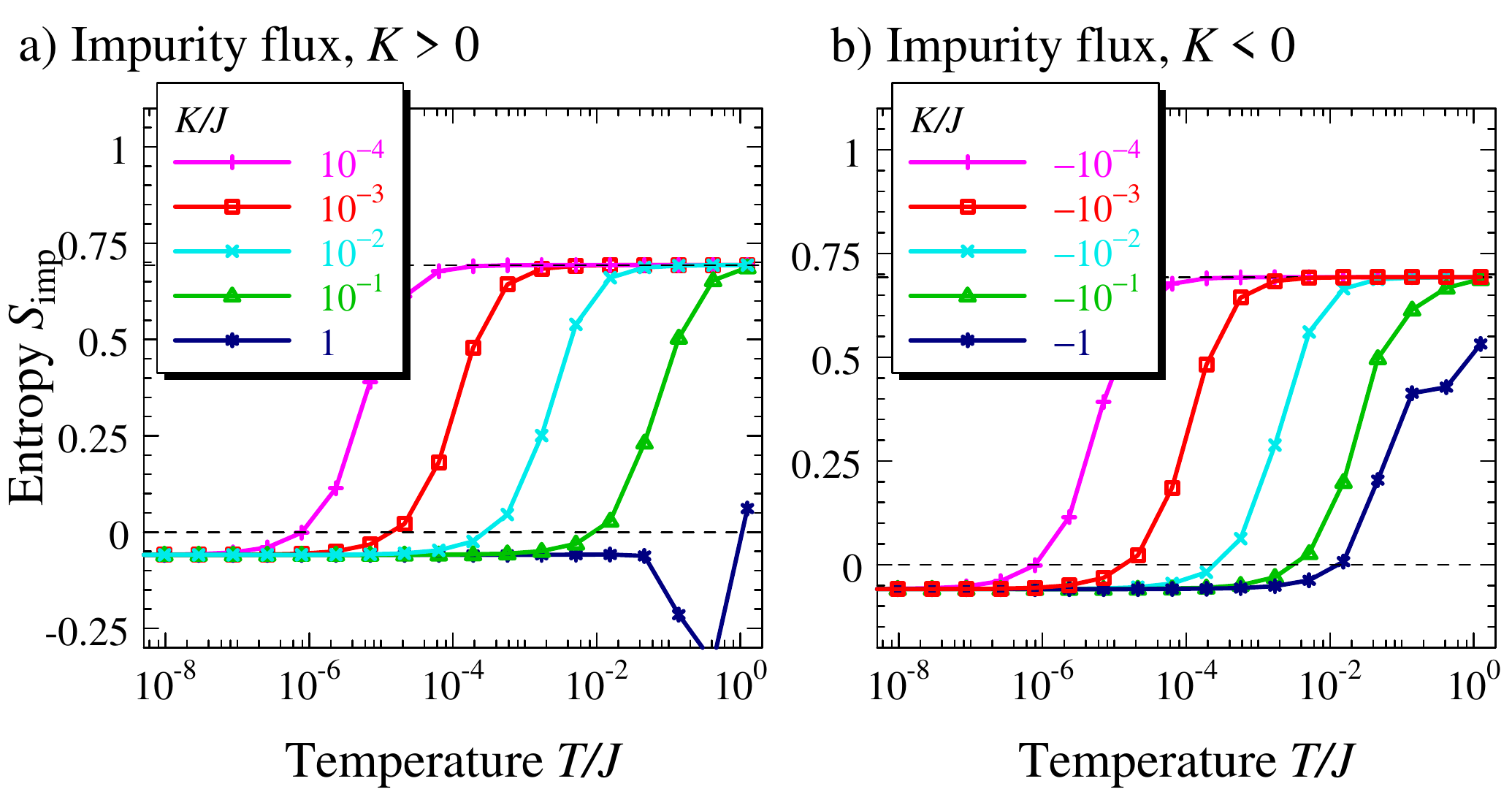}
\caption{\label{fig:s_flux1}
Impurity entropy as in Fig.~\ref{fig:s_noflux1}, but for the vacancy-flux case, with $J'=J$ and various $K$.
}
\end{center}
\end{figure}

\textit{NRG results: Impurity-flux case.}
We now turn to the case where the impurity plaquette is threaded by a $\Ztwo$ flux, $W_I=-1$, while
$W_p=+1$ otherwise. NRG results for the impurity entropy are shown in Fig.~\ref{fig:s_flux1}. We find that $\Simp \approx -0.06$ at low $T$ for all $K$.

For $J'\!=\!J$ there is a single crossover upon cooling, here from $\Simp\!=\!\ln 2$ to $-0.06$, with a crossover scale $T^\ast \propto |K|$ for both signs of $K$. The NRG levels identify the intermediate $\ln 2$ fixed point as LM, whereas the low-$T$ fixed points correspond to $K\to\infty$ and $K\to-\infty$, i.e. SVac and TVac$'$, respectively.
Notably, we also observe a clear TVac$\to$TVac$'$ crossover for small $J'$ and large negative $K$ \cite{suppl}: On lowering the temperature through $T^\ast \propto J'^2/J$, the entropy decreases from $\Simp \approx 1.04$ to $-0.06$.

\textit{Analytics: Flux-free case.}
Analytical considerations enable essentially a full understanding of the numerical findings. We start analyzing the vicinity of LM in the flux-free sector. Here, the Kondo spin is coupled to a Majorana bath at site 0 with a power-law density of states (DOS), $\rho(\w)=|\w|^r$ with $r=1$.
This problem is related to the extensively studied pseudogap Kondo model \cite{withoff,GBI,FV04,DEL} at particle--hole (p-h) symmetry; recall that on-site potentials are forbidden for Majorana fermions.
Importantly, the p-h symmetric pseudogap Kondo model exhibits no screening, even for strong antiferromagnetic Kondo coupling, because the relevant resonant-level fixed point is unstable for $r>1/2$ \cite{GBI,FV04,DEL}. This argument carries over to the present case, implying that SVac must be unstable in the flux-free case. In fact, the relevant perturbation to SVac has scaling dimension unity and initial value $J'^3/(JK^2)$; SVac is destabilized once this perturbation reaches the scale $K^2/J^2$ \cite{suppl}, explaining the numerically identified $T^\ast$.

Refs.~\onlinecite{dhochak10,dhochak15} argued that the flux-free sector displays a QPT between unscreened and screened phases. However, this is not the case -- the QPT is an artifact of weak-coupling RG \cite{FV04,suppl}, not observed in the NRG solution.

It is instructive to analyze $\Simp$ at the unstable SVac fixed point, where the Kondo spin and the host spin at site $0$ form a tightly bound singlet complex. Since site $0$ is then effectively cut out from the host \cite{willans10}, the three dangling gauge Majorana fermions $\hat{b}_1^x$, $\hat{b}_2^y$, $\hat{b}_3^z$, give a residual entropy $\tfrac{3}{2}\ln 2$. But $\Simp$ is the entropy \textit{relative} to that of the Kitaev host with no impurity (and therefore no vacancy), and so we must subtract the contribution of the Kitaev site $0$ to the host entropy to obtain $\Simp$. Site $0$ is coupled to three spinless electron channels, and can itself be viewed as a Majorana resonant-level model. The s-wave channel responsible for screening site $0$ has a power-law diverging DOS
$1/(\w\ln^2\w)$, and so its residual entropy \cite{GBI,grvac,fracsp} is $(-\tfrac{1}{2}\ln 2)$ (the factor of $\tfrac{1}{2}$ arising here because we are dealing with Majoranas).
Overall, $\Simp^{\rm SVac}=(\tfrac{3}{2}\ln 2)-(-\tfrac{1}{2}\ln 2) =\ln 4$ as found from NRG.

At TVac, the Kondo impurity and spin $0$ form an $S\!=\!1$ triplet which contributes an additional $\ln 3$ entropy, giving overall $\Simp^{\rm TVac} = \Simp^{\rm SVac} + \ln 3 = \ln 12$, again confirmed by NRG \cite{suppl}.




\textit{Analytics: Impurity-flux case.}
It is important to realize that the condition $W_I=-1$ induces a finite DOS at site $0$ when $J'\neq 0$ \cite{willans10}, and it imposes a threefold bath degeneracy because $W_I=-1$ can be achieved with a flux in any of the three plaquettes next to site $0$ (the fourth configuration with fluxes through all three plaquettes is higher in energy). This degeneracy is lifted by the Kondo impurity; hence $K$ couples two degenerate subsystems -- the Kondo spin and the bath -- resulting in an entropy quench and concomitant partial screening. In contrast to standard Kondo problems \cite{hewson,FV04}, the Kondo temperature is therefore $T^\ast \propto |K|$.

The $\Simp$ values arise as follows: Due to the subtraction of the impurity-free reference system, $\Simp^{\rm LM}=\ln 2$ despite the threefold degeneracy of the Kitaev host. However, $\Simp$ is lower by $\ln 3$ at the $K\neq0$ fixed points, due to the lifted flux degeneracy. As before, SVac and TVac  ($K\!=\!\pm\infty$) contain three dangling gauge Majoranas. Since there is no bath divergence in the impurity-flux case, we have $\Simp^{\rm SVac}=\tfrac{3}{2} \ln 2 - \ln 3 \approx -0.06$ and $\Simp^{\rm TVac}=\Simp^{\rm SVac}+\ln 3=\tfrac{3}{2}\ln 2$.
For ferromagnetic $K$, broken spin symmetry drives the flow to TVac$'$: the spin-triplet degeneracy is lifted such that $\Simp^{{\rm TVac}'}=\Simp^{\rm SVac}$.


\begin{figure}[t!]
\begin{center}
\includegraphics[width=8.7cm]{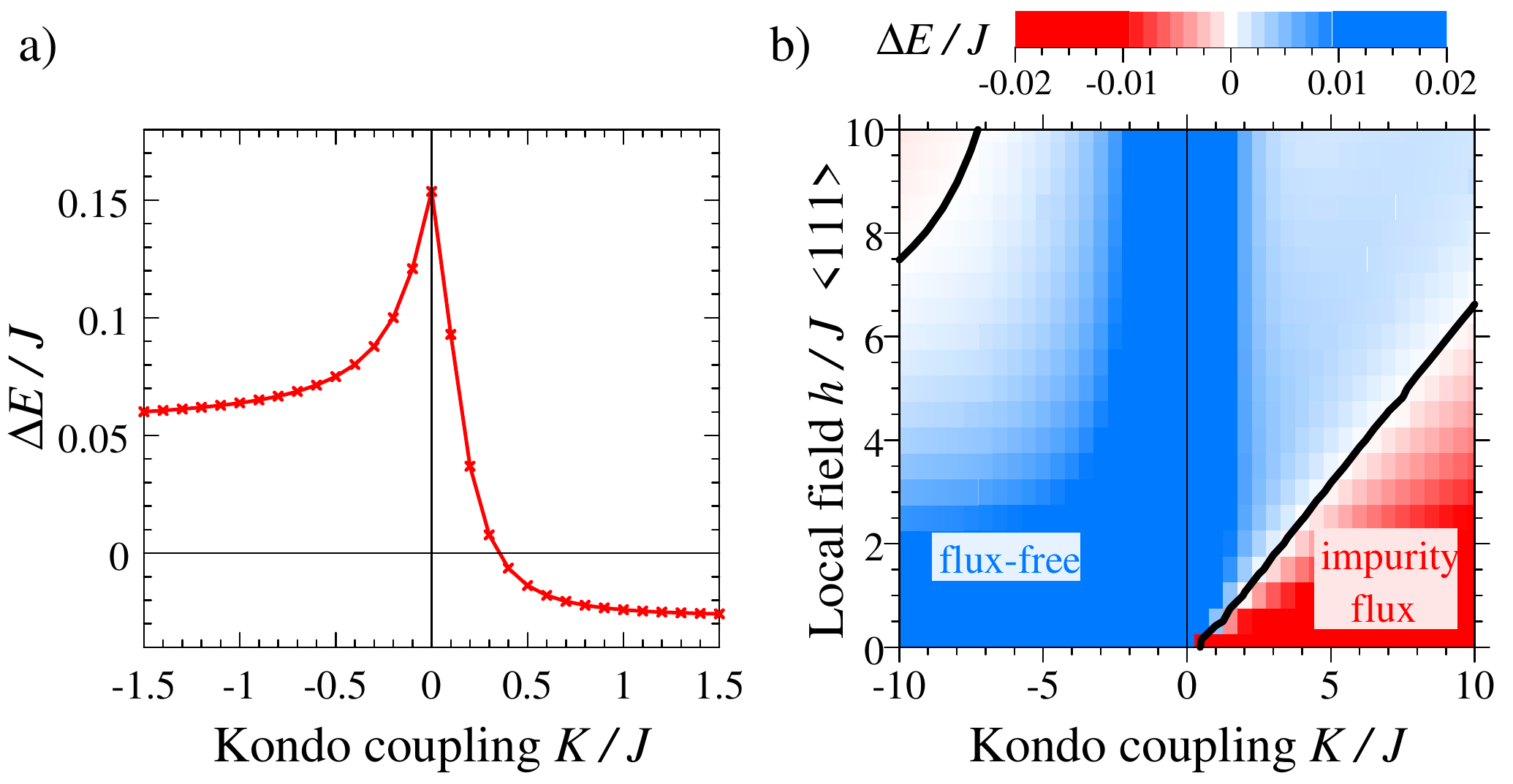}
\caption{\label{fig:energies}
Energy difference between impurity-flux and no-flux ground states, $\Delta E =  E_{\text{flux}}-E_{\text{noflux}}$, as function of $K/J$
(a) at $h=0$ and
(b) as a function of $h$, with the lines denoting phase boundaries.
}
\end{center}
\end{figure}


\textit{Field response.}
The response to a local magnetic field $h$ characterizes the fate of the Kondo spin. In the stable LM phase of the flux-free case, the Kondo spin is unscreened and the local susceptibility displays a low-$T$ Curie law, $\cloc(T) = \Cloc/T$, arising from the residual magnetic moment on the Kondo site, with $\Cloc\leq1$ (and $\Cloc=1$ only at $K=0$) \cite{sunityfoot}.

For the impurity-flux case we first note that the $K=\infty$ model is known to display a weakly singular response arising from the three dangling gauge Majorana fermions, $\chi \propto 1/\ln T$ \cite{willans10}. Rather surprisingly, we find that the SVac phase displays a low-$T$ Curie law for any $K<\infty$, but with a much reduced $\Cloc$. This Curie response arises from a subtle interplay of Majorana zero modes and the tightly bound singlet: Virtual triplet excitations couple pairs of zero modes, e.g., $\hat{b}_1^x$ and $\hat{b}_2^y$ acquire a coupling, producing an effective free moment along $z$ \cite{suppl}. Together, this results in partial screening. A similar Curie law is found in the TVac$'$ phase, but with $\Cloc$ of order unity because virtual excitations are less costly.

Beyond linear response, $h$ quenches the entropy contributions both from the residual moment and the localized Majorana zero modes \cite{suppl}.


\textit{Flux transition.}
Given that NRG calculations are performed in each flux sector, the global ground state is determined by comparing NRG ground-state energies. The energy difference $\Delta E$ between the impurity-flux and flux-free sectors is shown in Fig.~\ref{fig:energies}a.
We conclude that, at zero field, a first-order quantum phase transition occurs at $K_c\approx0.35J$ between a flux-free unscreened LM phase and a partially screened SVac phase with impurity flux. A local field $h$ drives multiple first-order transitions; for $h$ along the $\langle111\rangle$ direction flux binding can also occur for ferromagnetic $K$, Fig.~\ref{fig:energies}(b).

Finally, we note that for elevated temperatures, $T \gtrsim |\Delta E|$, the behavior is given by a thermal mixture of different flux sectors, as demonstrated for the plain Kitaev model in Ref.~\onlinecite{motome14}.


\textit{Finite impurity concentration.}
The physics at small but finite defect concentration is a rich subject. Due to the absence of extended spin correlations in the Kitaev model, residual moments do not mutually interact \cite{willans11}. Defect-induced magnetic order will therefore only emerge on taking into account bulk interactions beyond Kitaev \cite{jackeli09,Cha10,kee14} -- this is beyond the scope of the present work. However, the flux binding is a robust feature: The disordered flux arrangement for substitutional $S\!=\!0$ impurities (effectively realized for $K>K_c$) will strongly scatter matter Majoranas and dramatically decrease the magnetic low-$T$ thermal conductivity; this does not apply to $S\!=\!1$ impurities which do not bind fluxes.


\textit{Conclusions.}
We have solved the Kondo problem for the gapless Kitaev model using NRG -- this represents the first numerically exact solution for a quantum impurity coupled to a spin liquid. We have determined the phase diagram and characterized the RG fixed points in terms of localized Majorana zero modes. The case of a $S\!=\!1$ substitutional spin is encompassed by our solution at large ferromagnetic $K$; it behaves like a free moment with Curie susceptibility.
Our approach can be extended to Kitaev models on other lattices \cite{yao07,baskaran09,kiv1,hermanns} and more complicated impurity problems.
Experimental realizations using magnetic adatoms on layers of $\alpha$-RuCl$_3$ \cite{rucl} or related materials appear possible.


\textit{Acknowledgements.}
We thank D. P. Arovas, P. P. Baruselli, L. Fritz, T. Meng, and S. Rachel for instructive discussions and for collaborations on related work.
M.V. and F.Z. thank the DFG for support via SFB 1143 and GRK 1621 and the Helmholtz association via VI-521. A.K.M. acknowledges the D-ITP consortium, a program of the Netherlands Organisation for Scientific Research (NWO) that is funded by the Dutch Ministry of Education, Culture and Science (OCW).



\begin{thebibliography}{99}
\bibitem{sunityfoot}
We define the Kondo spin as Pauli matrix without factor $1/2$, as done for the bulk Kitaev spins. As a result, the Curie constant of a free spin is $C=1$ (instead of $1/4$).
\end{thebibliography}
\end{document}


\title{
Supplemental material: \\
Kondo impurities in the Kitaev spin liquid: \\
Numerical renormalization group solution and gauge-flux-driven screening
}

\author{Matthias Vojta}
\affiliation{Institut f\"ur Theoretische Physik, Technische Universit\"at Dresden, 01062 Dresden, Germany}
\author{Andrew K. Mitchell}
\affiliation{Institute for Theoretical Physics, Utrecht University, 3584 CE Utrecht, The Netherlands}
\author{Fabian Zschocke}
\affiliation{Institut f\"ur Theoretische Physik, Technische Universit\"at Dresden, 01062 Dresden, Germany}


\date{\today}

\pacs{}

\maketitle

\section{Bulk Kitaev model}

In this section we summarize aspects of the Majorana representation of the Kitaev model which are relevant to the solution of the Kondo model via Wilson's Numerical Renormalization Group (NRG) technique.

We assume a honeycomb lattice with $N$ unit cells and $2N$ spins. In order to cover the cases with modified couplings $J'$ near the impurity, we generalize the Kitaev model to spatially inhomogeneous couplings:
\begin{equation}
\label{hk}
\HK =
-\sum_{\langle ij\rangle_x} J_{ij}^x \hat{\sigma}_i^x \hat{\sigma}_j^x
-\sum_{\langle ij\rangle_y} J_{ij}^y \hat{\sigma}_i^y \hat{\sigma}_j^y
-\sum_{\langle ij\rangle_z} J_{ij}^z \hat{\sigma}_i^z \hat{\sigma}_j^z
\end{equation}

\subsection{Majorana representation}

Following Kitaev's solution,\cite{kitaev06} we introduce four (real) Majorana fermions per site: $\hat{b}^x$, $\hat{b}^y$, $\hat{b}^z$, and $\hat{c}$ .
Defining $\hat{\sigma}_i^\alpha=\ii \hat{b}_i^\alpha \hat{c}_i$ the Hamiltonian in Eq.~\eqref{hk} can be mapped to
\begin{equation}
\label{hm}
\HMh = \ii\sum_{\langle ij\rangle}J^\alpha_{{ij}} \hat{u}_{ij}\hat{c}_i \hat{c}_j,
\end{equation}
where $\hat{u}_{ij}\equiv i\hat{b}_i^{\alpha_{ij}}\hat{b}_j^{\alpha_{ij}}$ and $\hat{u}_{ij}=-\hat{u}_{ji}$. We follow the convention that, when specifying $\hat{u}_{ij}$, $i$ is located on sublattice $A$.
The operators $\hat{u}_{ij}$ commute with each other and the Hamiltonian $\HM$
and have eigenvalues of $u_{ij}=\pm 1$. A given set $\{u_{ij}\}$ reduces the Hamiltonian to a bilinear in the $\hat{c}$ (``matter'') Majorana operators:
\begin{equation}
\label{hmflux}
\HM = \frac{\ii}{2}\left(\hat{c}^T_A \, \hat{c}^T_B\right) \begin{pmatrix}
0 & M \\
-M^T & 0
\end{pmatrix}
\begin{pmatrix}
\hat{c}_A \\
\hat{c}_B
\end{pmatrix}.
\end{equation}
Here $\hat{c}_{A(B)}$ is a vector of length $N$ of Majorana operators on the $A(B)$ sublattice, and $M$ is an $N \times N$ matrix with elements $M_{ij}=J^\alpha_{ij} u_{ij}$, reflecting the coupling of the matter Majorana fermions to the $\Ztwo$ gauge field encoded by $u_{ij}$.

The eigenmodes of $\HM$ can be found via singular-value decomposition of $M$, $M=USV^T$, where $U$ and $V$ are $N\times N$ orthogonal matrices, and $S$ is an $N\times N$ diagonal matrix containing the non-negative singular values of $M$. We define new Majorana operators according to
\begin{equation}
\begin{split}
\label{uvtrans}
(\hat{b}'_{1},\ldots,\hat{b}'_{N})&=(\hat{c}_{A,1},\ldots,\hat{c}_{A,N}) U \,,\\
(\hat{b}''_{1},\ldots,\hat{b}''_{N})&=(\hat{c}_{B,1},\ldots, \hat{c}_{B,N}) V \,.
\end{split}
\end{equation}
We may combine the transformation matrices $U$ and $V$ into a matrix $Q^u$,
\begin{equation}
\label{traf}
Q^u= \begin{pmatrix}
0 & U \\
V & 0
\end{pmatrix},
\end{equation}
which is equivalent to $Q^u$ defined in Eq.~(4) of Ref.~\onlinecite{loss} after re-ordering of both rows and columns.

For a given set of $\{u_{ij}\}$ the Hamiltonian now has the form $\HM = \ii \sum^N_{m=1} \e_m \hat{b}'_{m}\hat{b}''_{m}$, where $\e_m \geq 0$ are the singular values of $M$.
It is convenient to combine the Majorana operators $\hat{b}'$, $\hat{b}''$ into canonical fermions according to
\begin{equation}
\label{adef}
\hat{a}_m=\frac{1}{2}(\hat{b}'_{m} + \ii \hat{b}''_{m})\,.
\end{equation}
This eventually gives
\begin{equation}
\label{hmsvd}
\HM = \sum^N_{m=1} \e_m ( 2 \hat{a}^\dagger_{m}\hat{a}_{m}-1)
\end{equation}
with the ground-state energy $E_0=-\sum_m \e_m$.

\subsection{Flux degrees of freedom}

For every closed loop $C$ of the lattice, there exists a conserved quantity described by an observable $\hat{W}_C$.\cite{kitaev06} For a loop $C$ containing $L$ sites labeled $\{1,2,...,L\}$, this observable is
\begin{equation}
\hat{W}_C=\hat{\sigma}_1^{\alpha_{1,2}}\hat{\sigma}_2^{\alpha_{1,2}}\hat{\sigma}_2^{\alpha_{2,3}}\hat{\sigma}_3^{\alpha_{2,3}}\dots \hat{\sigma}_L^{\alpha_{L,1}} \hat{\sigma}_1^{\alpha_{L,1}},
\end{equation}
with eigenvalues $W_C=\pm1$, each corresponding to a $\Ztwo$ flux. Loop operators for the fluxes through each elementary plaquette of the lattice are introduced as
\begin{equation}
 \hat{W}_p = \hat{\sigma}_1^x \hat{\sigma}_2^y \hat{\sigma}_3^z \hat{\sigma}_4^x \hat{\sigma}_5^y \hat{\sigma}_6^z.
\end{equation}
For periodic boundary conditions there are, in addition, two ``topological'' loop operators $W_{1,2}$ that wrap around the torus.

In the Majorana representation, the loop (or flux) operators $\hat{W}$ can be expressed through the bond variables $\hat{u}_{ij}$; the same holds for their eigenvalues. For instance, the plaquette fluxes take the form
\begin{equation}
W_p = u_{21}u_{23}u_{43}u_{45}u_{65}u_{61}\,.
\end{equation}
As a consequence of gauge invariance, the fermion spectrum $\e_m$ depends on the $u_{ij}$ only through the values of the fluxes $W_C$.

\subsection{Physical states and boundary conditions}

For the Kondo problem we will be interested in the thermodynamic limit of the host; hence the type of boundary conditions should be irrelevant. In order to avoid the subtleties concerning the selection of physical states discussed in Refs.~\onlinecite{loss} and \onlinecite{zschocke15}, we assume open boundaries before taking the thermodynamic limit.

We note that the full energy dependence of the NRG bath must be calculated from finite-size Kitaev systems. To minimize finite-size effects we employ periodic boundary conditions and take the thermodynamic limit by extrapolating the bath spectrum to low energies, see below.


\section{NRG formulation for the Kitaev Kondo model}

Wilson's NRG can be applied to problems where a (possibly complex) impurity is coupled to a bath of {\it non-interacting} canonical particles.\cite{nrg,nrg_rev} The latter applies to the Kitaev host in each individual flux sector. However, the $\Ztwo$ fluxes in the three elementary plaquettes next to the Kondo impurity are not conserved under the action of the Kondo term, $\sum_\alpha K^\alpha \hat{S}^\alpha \hat{\sigma}_0^\alpha$, i.e., they become dynamical. Their product, corresponding to the flux through the impurity plaquette (see Fig.~{\figsetup} of the main paper), remains conserved, however. Hence, the plaquette fluxes next to the Kondo site must be included into a generalized ``impurity'' within the NRG treatment.

The non-conservation of the fluxes near the impurity also implies that the NRG bath should include all $\hat{c}$ (matter) Majorana fermions \textit{except} $\hat{c}_0$ residing at site $0$. After removing $\hat{c}_0$ with its links from the Kitaev host, the bath displays a delocalized Majorana zero mode, akin to the single-vacancy problem in graphene. It is convenient to exclude this zero mode from the bath as well, in order to have an even number of matter Majorana fermions in the definitions of the NRG bath and the NRG impurity.

\subsection{NRG treatment: Bath}

The Hilbert space of the NRG bath is that of a hopping problem of Majorana fermions on a honeycomb lattice with one ``missing'' site and a fixed configuration of $\Ztwo$ fluxes. It is governed by the Hamiltonian
\begin{equation}
\Hbath = \HM |_{J_{0j}=0}
\end{equation}
with $\HM$ in Eq.~\eqref{hmflux}. As explained above, the bath can be transformed to non-interacting canonical fermions. For any flux configuration, the bath will have one fermion zero mode.
This becomes most transparent by considering a Majorana hopping problem where the links to site $0$ are switched off, such that site $0$ is dangling. Then, one of the excitation energies $\e_m$ in Eq.~\eqref{hmsvd} vanishes, and we denote the corresponding canonical fermion by $\hat{a}_0$. This consists of two $\hat{c}$ Majoranas, one on sublattice A which is the dangling Majorana fermion at site $0$ and one on sublattice B which is the (delocalized) vacancy-induced zero mode. Both $\hat{c}_0$ and the delocalized zero mode will be included into the NRG impurity.

Provided that the $\Zthree$ rotation symmetry w.r.t. the impurity site is preserved, the bath modes may be decomposed into angular-momentum channels (more precisely: irreducible representations of the point group). As the impurity couples to the three bath sites next to site $0$ -- sites $1,2,3$ in Fig.~{\figsetup} of the main paper -- we will decompose the bath propagator at these three sites into the three relevant angular-momentum channels, dubbed $s$ and $p_\pm$ in the following.

While a full solution of the Kondo problem would require to consider all flux sectors, we will restrict our attention to low energies and temperatures. While the ground-state flux sector is a-priori not known, we make use of two known facts:
(i) The plain Kitaev model has its ground state in the flux-free sector;\cite{kitaev06} this will therefore apply to the Kondo model at small $K$.
(ii) For the Kitaev model with a single vacancy the ground state is in the sector with a $\Ztwo$ flux attached to the vacancy plaquette, but all other plaquettes flux-free.\cite{willans10,willans11} This will carry over to the Kondo model at large antiferromagnetic $K$.
Hence, we will restrict our attention to these two flux sectors, dubbed ``flux-free'' and ``impurity flux'' in the following.

\subsection{NRG treatment: Impurity}

As noted above, the Hilbert space of the NRG impurity has to include the fluxes through the plaquettes next to site $0$  -- this is equivalent to including the dangling gauge Majorana fermions $\hat{b}_1^x$, $\hat{b}_2^y$, $\hat{b}_3^z$ into the impurity, together with the $\hat{b}_0^\alpha$. This yields six gauge Majorana fermions with one constraint, the flux through the impurity plaquette, resulting in an impurity flux/gauge Hilbert space of four states. (The same counting is trivially obtained from having three plaquette fluxes with one constraint.)
Thus the impurity Hilbert space consists of 16 states: two states of the Kondo spin times four gauge/flux states times two matter states of the canonical fermion $\hat{a}_0$; recall that $\hat{a}_0$ contains $\hat{c}_0$ and the delocalized zero mode of the bath.

$\HKo$ from Eq.~{\eqhko} acts exclusively in this impurity Hilbert space. Its Majorana representation reads
\begin{align}
\HKo &= \ii \sum_\alpha K^\alpha \hat{S}^\alpha \hat{b}_0^\alpha \hat{c}_0 + \sum_\alpha h^\alpha S^\alpha
\end{align}
To proceed, we choose a basis in the impurity Hilbert space. For the Kondo spin we work in the basis of $\hat{S}^z$ eigenstates. For the gauge states, we take the eigenstates of $\hat{u}_{01}=\ii\hat{b}_0^x\hat{b}_1^x$ and $\hat{u}_{02}=\ii\hat{b}_0^y\hat{b}_2^y$, while the value of $u_{03}$ is kept fixed by choosing a suitable gauge. For the matter fermion, we take the occupation-number eigenstates of $\hat{a}_0$.
For the evaluation of matrix elements the following pieces of information are needed: (i) $\hat{c}_0$ appearing in $\HKo$ is $\hat{c}_0 = \hat{a}^\dagger_0 + \hat{a}_0$. (ii) The action of $\hat{b}_0^z$ which changes the value of $u_{03}$ is supplemented by acting with $\hat{D}_i=\hat{b}_i^x\hat{b}_i^y\hat{b}_i^z\hat{c}_i$ on site $i=0$ -- this operator can be thought of as a gauge transformation and has eigenvalue $+1$ when acting on physical states.\cite{kitaev06}

We order the basis states as follows:
$|\uparrow,00,0\rangle$,
$|\uparrow,10,0\rangle$,
$|\uparrow,01,0\rangle$,
$|\uparrow,11,0\rangle$,
$|\uparrow,00,1\rangle$,
$|\uparrow,10,1\rangle$,
$|\uparrow,01,1\rangle,$
$|\uparrow,11,1\rangle$,
$|\downarrow,00,0\rangle$, \ldots,
$|\downarrow,11,1\rangle$.
Then, the impurity piece of the Hamiltonian takes the matrix form:
\begin{widetext}
\begin{align}
&\Himp = \\
&\left(
\begin{smallmatrix}
h^z	& 0        & 0        & -\ii K^z    & 0           & 0         & 0        & 0        & h^x+\ii h^y & 0        & 0        & 0        & 0        & \ii K^x     & -K^y     & 0        \\
0 	& h^z      & \ii K^z     & 0        & 0           & 0         & 0        & 0        & 0        & h^x+\ii h^y & 0        & 0        & -\ii K^x    & 0        & 0        & -K^y     \\
0 	& -\ii K^z    & h^z      & 0        & 0           & 0         & 0        & 0        & 0        & 0        & h^x+\ii h^y & 0        & K^y      & 0        & 0        & -\ii K^x    \\
\ii K^z 	& 0        & 0        & h^z      & 0           & 0         & 0        & 0        & 0        & 0        & 0        & h^x+\ii h^y & 0        & K^y      & \ii K^x     & 0        \\
0 	& 0        & 0        & 0 	 & h^z         & 0         & 0        & -\ii K^z    & 0        & \ii K^x     & -K^y     & 0        & h^x+\ii h^y & 0        & 0        & 0        \\
0 	& 0        & 0        & 0 	 & 0           & h^z       & \ii K^z     & 0        & -iK^x    & 0        & 0        & -K^y     & 0        & h^x+\ii h^y & 0        & 0        \\
0 	& 0        & 0        & 0 	 & 0           & -\ii K^z     & h^z      & 0        & K^y      & 0        & 0        & -\ii K^x    & 0        & 0        & h^x+\ii h^y & 0        \\
0 	& 0        & 0        & 0 	 & \ii K^z        & 0         & 0        & h^z      & 0        & K^y      & \ii K^x     & 0        & 0        & 0        & 0        & h^x+\ii h^y \\
h^x-\ii h^y& 0        & 0        & 0 	 & 0           & \ii K^x      & K^y      & 0        & -h^z     & 0        & 0        & \ii K^z     & 0        & 0        & 0        & 0        \\
0 	& h^x-\ii h^y & 0        & 0        & -\ii K^x       & 0         & 0        & K^y      & 0        &-h^z      & -\ii K^z    & 0        & 0        & 0        & 0        & 0        \\
0 	& 0        & h^x-\ii h^y & 0        & -K^y        & 0         & 0        & -\ii K^x    & 0        & \ii K^z     & -h^z     & 0        & 0        & 0        & 0        & 0        \\
0 	& 0        & 0        & h^x- \ii h^y & 0           & -K^y      & \ii K^x     & 0        & -\ii K^z    & 0        & 0        & -h^z     & 0        & 0        & 0        & 0        \\
0 	& \ii K^x     & K^y      & 0        & h^x-ih^y    & 0         & 0        & 0        & 0        & 0        & 0        & 0        & -h^z     & 0        & 0        & \ii K^z     \\
-\ii K^x 	& 0        & 0        & K^y      & 0           & h^x-\ii h^y  & 0        & 0        & 0        & 0        & 0        & 0        & 0        & -h^z     & -\ii K^z    & 0        \\
-K^y 	& 0        & 0        & -\ii K^x    & 0           & 0         & h^x-\ii h^y & 0        & 0        & 0        & 0        & 0        & 0        & \ii K^z     & -h^z     & 0        \\
0 	& -K^y     & \ii K^x     & 0        & 0           & 0         & 0        & h^x-\ii h^y & 0        & 0        & 0        & 0        & -\ii K^z    & 0        & 0        & -h^z     \\
\end{smallmatrix}
\right)
\notag
\end{align}
\end{widetext}
which can be directly implemented into the NRG code.\cite{sunityfoot}

\subsection{NRG treatment: Hybridization}

The coupling between impurity and bath is captured by the hybridization piece,
\begin{equation}
\Hhyb = \ii \sum_{i=1}^{3} J'^i \hat{u}_{0i} \hat{c}_0 \hat{c}_i
\end{equation}
where $\hat{u}_{0i} = \ii \hat{b}_0^i \hat{b}_i^i$ with $i$ taking values $1\equiv x$, $2\equiv y$, and $3\equiv z$. The $\hat{c}_{1,2,3}$ Majorana fermions -- all living on the B sublattice -- are related to the bath eigenmodes according to
\begin{align}
\label{c123rep}
-\ii c_i = \sum_n V_{in} (\hat{a}_n^\dagger - \hat{a}_n) ~~~(i=1,2,3)
\end{align}
with $V$ being the real orthogonal matrix from Eq.~\eqref{uvtrans}.

To facilitate an angular-momentum decomposition of the bath we introduce linear combinations of the matter Majoranas at sites $1,2,3$ according to
\begin{equation}
\hat{d}_m = \frac{1}{\sqrt{3}}\sum_{i=1}^{3} e^{\ii (i-1) m 2\pi/3} \hat{c}_i
\end{equation}
with $m=0,\pm 1$ corresponding to angular-momentum channels. Note that $d_0=d_0^\dagger$, but $d_{\pm 1}=d_{\mp 1}^\dagger$.
Using Eq.~\eqref{c123rep} we have
\begin{align}
-\ii \hat{d}_m &= \tilde{V}_{m0} (\hat{a}_0^\dagger - \hat{a}_0) + \sum'_n \tilde{V}_{mn} (\hat{a}_n^\dagger - \hat{a}_n) \notag\\
&= \tilde{V}_{m0} (\hat{a}_0^\dagger - \hat{a}_0) -\ii \hat{d}'_m
\end{align}
where $\tilde{V}_{mn} = \sum_i e^{\ii (i-1) m 2\pi/3} V_{in} /\sqrt{3}$, and we have split off the vacancy-induced zero mode of the bath which is excluded in the sum $\sum'_n$. Note that the $\tilde{V}_{\pm1n}$ are no longer real, and we have $\tilde{V}^\ast_{1n} = \tilde{V}_{-1n}$.
Further, rotation symmetry implies $\tilde{V}_{m0}=0$ for $m=\pm1$.

We now introduce spinless canonical fermions $\hat{\Psi}_m$ to represent the bath degrees of freedom (excluding the zero mode) at sites 1,2,3 in the relevant angular-momentum channels:
\begin{align}
\beta_m \hat{\Psi}_m = \sum'_n \tilde{V}_{mn} \hat{a}_n
\end{align}
such that
\begin{align}
-\ii \hat{d}'_m = \beta_{-m}\hat{\Psi}_{-m}^\dagger-\beta_m\hat{\Psi}_m.
\end{align}
Here $\beta_m$ is a real number accounting for the proper normalization of $\hat{\Psi}_m$ which is required due to the missing zero mode. Specifically, $\beta_0^2 = 1- \tilde{V}_{00}^2$ and  $\beta_{\pm1} = 1$ due to $\tilde{V}_{\pm10}=0$.

With these ingredients we can re-write the hybridization piece as follows
\begin{equation}
\Hhyb = X_0 \tilde{V}_{00} + \sum_{m=-1}^1 (Y_m \beta_m \hat{\Psi}_m + h.c.)
\label{hyb2}
\end{equation}
where $X_0$ and $Y_m$ describe the hybridization of site $0$ with the zero-energy and finite-energy modes of the bath, respectively. Both $X_0$ and $Y_m$ are matrices in the 16-state impurity Hilbert space. Given that $X_0$ and $Y_m$ do not act on the Kondo spin, we specify them in the reduced Hilbert space excluding the Kondo spin. Adopting the ordering from above, i.e.,
$|00,0\rangle$,
$|10,0\rangle$, \ldots
$|11,1\rangle$,
the matrices in the $s$-wave channel are
\begin{widetext}
\begin{align}
X_0 &=
\frac{1}{\sqrt{3}}
\left(
\begin{smallmatrix}
J'_x+J'_y+J'_z & 0 & 0 & 0 & 0 & 0 & 0 & 0\\
0 & -J'_x+J'_y+J'_z & 0& 0 & 0 & 0 & 0 & 0 \\
0 & 0 & J'_x-J'_y+J'_z & 0 & 0 & 0 & 0 & 0  \\
0 & 0 & 0 & -J'_x-J'_y+J'_z & 0 & 0 &  0 & 0  \\
0 & 0 & 0 & 0 & -J'_x-J'_y-J'_z & 0 & 0 & 0  \\
0 & 0 & 0 & 0 & 0 & J'_x-J'_y-J'_z & 0 & 0  \\
0 & 0 & 0 & 0 & 0 & 0 & -J'_x+J'_y-J'_z & 0  \\
0 & 0 & 0 & 0 & 0 & 0 & 0 & J'_x+J'_y-J'_z  \\
\end{smallmatrix}
\right),
\\
Y_0 &=
\frac{1}{\sqrt{3}}
\left(
\begin{smallmatrix}
0 & 0 & 0 & 0 & J'_x+J'_y+J'_z & 0 & 0 & 0  \\
0 & 0 & 0 & 0 & 0 & J'_x-J'_y-J'_z & 0 & 0  \\
0 & 0 & 0 & 0 & 0 & 0 &  -J'_x+J'_y-J'_z & 0 \\
0 & 0 & 0 & 0 & 0 & 0 & 0 & -J'_x-J'_y+J'_z  \\
J'_x+J'_y+J'_z & 0 & 0 & 0 & 0 & 0 & 0 & 0 \\
0 &  J'_x-J'_y-J'_z & 0 & 0 & 0 & 0 & 0 & 0 \\
0 & 0 & -J'_x+J'_y-J'_z & 0 & 0 & 0 & 0 & 0 \\
0 & 0 & 0 & - J'_x-J'_y+J'_z & 0 & 0 & 0 & 0 \\
\end{smallmatrix}
\right).
\end{align}
\end{widetext}
The matrices in the $p$-wave channels are obtained by the replacements $J'_y \rightarrow J'_y e^{\pm \ii 2\pi /3}$ and $J'_z \rightarrow J'_z e^{\mp \ii 2\pi /3}$, such that $Y\ast_m = Y_{-m}$.
Hermiticity is ensured by noting that $(Y_m \Psi_m)^\dagger = \Psi_m^\dagger Y^\ast_m = -Y^\ast_m \Psi^\dagger_m = -Y_{-m} \Psi^\dagger_m$ where the fermionic character of $Y_m$ is taken into account.

In terms of NRG implementation, the coupling between impurity and bath is thus given by the second term in Eq.~\eqref{hyb2}, i.e., $\sum_m (Y_m \beta_m \hat{\Psi}_m + h.c.)$. This form implies that there is no particle-number conservation (but its parity is conserved).

In contrast, the first term of Eq.~\eqref{hyb2} acts within the Hilbert space of the NRG impurity only, i.e., needs to be added to $\Himp$. Its prefactor $\tilde{V}_{00}$ describes the amplitude of the vacancy-induced zero mode at the sites $1,2,3$. In the flux-free case, this has been studied in the context of graphene: $\tilde{V}_{00}=0$ in the infinite-system limit, but it is suppressed only logarithmically with system size.\cite{neto_rmp}
As our focus is on the thermodynamic limit of the bath, we take $\tilde{V}_{00}=0$; we have checked that this also applies in the vacancy-flux case.


\subsection{Bath propagators}

The NRG bath consists of three reservoirs of spinless fermions for the three angular-momentum channels. The properly normalized bath densities of states (DOS) corresponding to $\Psi$ in the three channels are
\begin{equation}
\rho_m(\w) = (1/\beta_m^2) \sum_n |\tilde{V}_{mn}|^2 \delta(\w-2\w_n),
\end{equation}
with the factor of two in the energy argument from Eq.~\eqref{hmsvd}. Note that $\tilde{V}_{mn}$ is dimensionless, and the ``hybridization strength'' is encoded in $Y_m$. In general, the densities of states are non-zero for positive energies only, as the singular values $\e_m$ are non-negative and the zero mode has been integrated into the impurity.

The specific $\rho_m(\w)$, being input for the NRG algorithm, must be obtained numerically in general, i.e., from finite-size simulations of Eq.~\eqref{hmflux}. This needs to be done separately for each flux configuration of the bath. As noted above, we will focus on the flux-free configuration and the one with a flux in the impurity plaquette; all other configurations are expected to lead to higher-energy states. We recall that the angular-momentum decomposition of the bath requires a flux configuration which preserves $\Zthree$ symmetry, this symmetry also ensures $\rho_{1} = \rho_{-1}$.
We finally mention that the $\hat{c}$ propagators are gauge-dependent, i.e., the NRG calculation is done in a fixed $\Ztwo$ gauge, suitably chosen to preserve the $\Zthree$ symmetry of the bath. As shown in Ref.~\onlinecite{kitaev06}, gauge fixing does not influence physical observables.

\begin{figure}[t]
\begin{center}
\includegraphics[width=85mm]{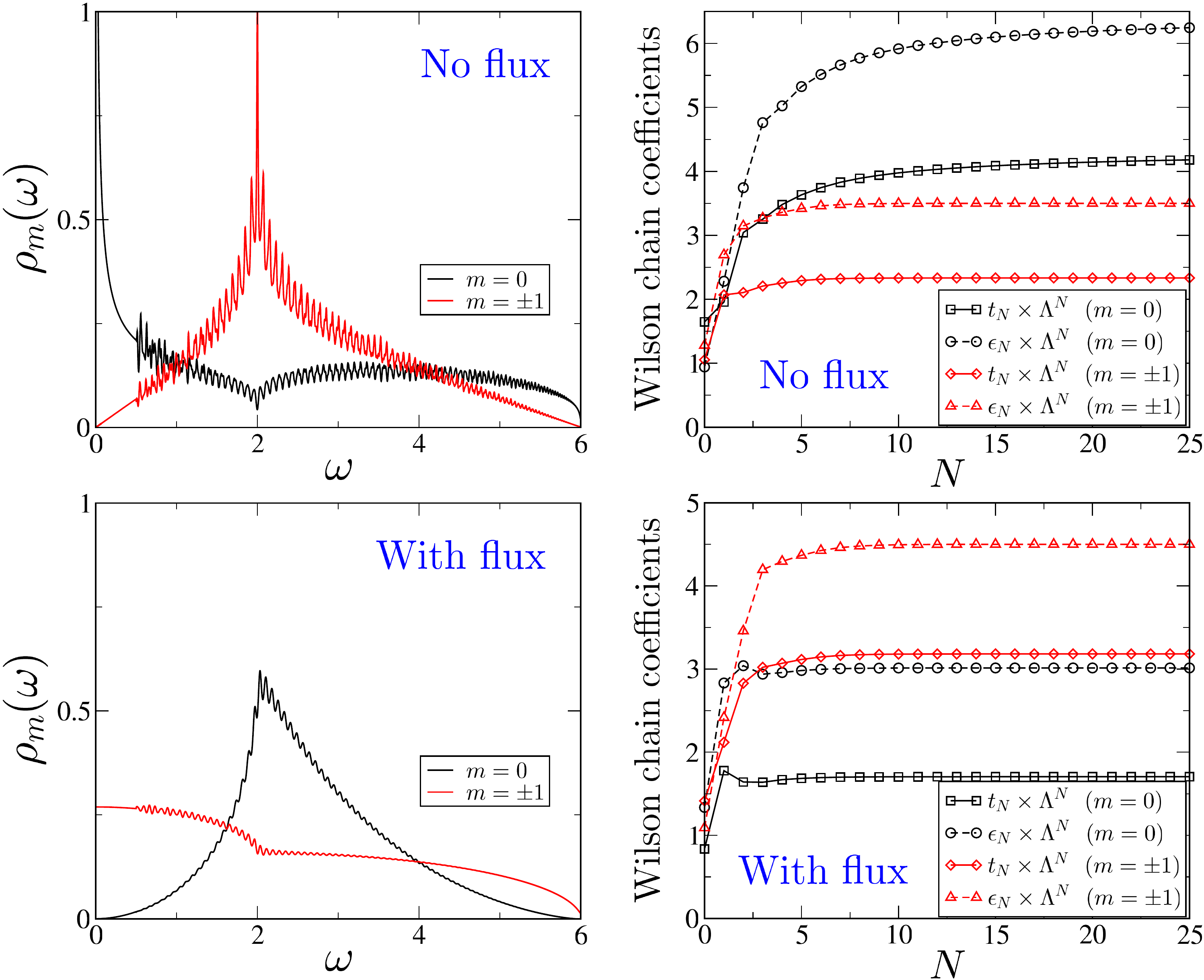}
\caption{\label{fig:wc}
Left panels: Density of states $\rho_m(\omega)$ vs $\omega$. Right panels: corresponding rescaled Wilson chain coefficients $t_N\times \Lambda^N$ and $\epsilon_N \times \Lambda^N$ vs NRG iteration number $N$. Upper and lower panels show the flux-free and impurity-flux cases, respectively, with black and red lines for $m=0$ and $m=\pm 1$ angular-momentum channels. The conduction electron band, of width $D=6J$, is discretized logarithmically using $\Lambda=2$. $\rho_m(\omega)$ is extrapolated to exponentially low energies using the identified asymptotic behavior, as required for determination of Wilson-chain coefficients.
}
\end{center}
\end{figure}

\subsubsection{Flux-free case}

In the flux-free case with isotropic hopping, the low-energy form of the honeycomb-lattice propagators is known analytically. This can be combined with a standard T-matrix calculation to obtain the low-energy asymptotic behavior of the NRG DOS in the three angular-momentum channels. For the $s$-wave ($m=0$) channel we find
\begin{equation}
\label{rho0}
\rho_0 (\w)= \frac{1}{J\pi} \left(\frac{2\pi^2}{\sqrt 3\omega(\pi^2+4[\ln(\omega/6)]^2)} \right)
\end{equation}
while in the $p$-wave channels
\begin{equation}
\rho_{\pm1} (\w)= \frac{1}{J\pi} \left(\frac{3\omega}{4\sqrt 3}-\frac{\omega^3}{48\sqrt 3 } \right).
\end{equation}
We recall that the local DOS of the unperturbed system is $\rho(\w) \propto \w$; cutting out site $0$ turns this into a divergence in the $s$-wave channel, $\rho_0(\w) \propto 1/\w$ with logarithmic corrections.

The numerical results for $\rho_m$ are in Fig.~\ref{fig:wc}. They have been obtained from a system with $N=160^2$ unit cells; the wiggles are effects of finite system size combined with Lorentzian broadening of $\gamma/J=0.01$. To generate the Wilson-chain coefficients, we use the numerical DOS for $\w/J>0.5$ and the asymptotic forms for smaller $\w$.

\subsubsection{Impurity-flux case}

In this case, no analytical progress can be made. For the finite-system numerics we have employed periodic boundary conditions and have placed two fluxes into the system, one in the impurity plaquette and one at the largest possible distance to the first. A $\Zthree$-symmetric configuration of $u_{ij}$ is most efficiently achieved having three strings of $u=-1$ bonds connecting the two fluxes; this implies the existence of a torus flux in addition. This is permissible, since its effect on local observables vanishes in the thermodynamic limit.

Numerical results for $\rho_m$ are in Fig.~\ref{fig:wc}(c). They appear consistent with the low-energy asymptotics $\rho_0 (\w) \propto \w^2$ and $\rho_{\pm 1} (\w) \to {\rm const.}$ -- we have employed such fitting functions to extrapolate $\rho(\w)$ down to zero energy. We emphasize that the vacancy flux qualitatively changes the bath propagators, as noted earlier;\cite{willans11} in particular it renders the local DOS at site $0$ (in the absence of the Kondo spin) finite. This is important for the stability of the SVac fixed point of the Kondo problem.

\subsection{NRG implementation}

In light of the above considerations, the NRG setup consists of a complex ``impurity'' comprising the Kondo spin-1/2 impurity itself, but including also the Kitaev spin at site $0$ and the surrounding gauge Majoranas. The hybridization term couples this $16\times 16$ impurity subsystem to the bath, which is described by a $3\times 3$ matrix propagator. The bath is cast into the form of three Wilson chains, corresponding to s- and p-wave channels. The nontrivial energy dependence of the hybridization is encoded in the Wilson chain coefficients, which are computed numerically using the Lanczos algorithm.\cite{nrg_rev} Since the hybridization term involves Majorana operators, the NRG calculation necessarily requires the use of complex numbers. Furthermore, there is no spin or particle number conservation, meaning that the NRG Hamiltonian cannot be block diagonalized. In practice, this limits the number of states, $N_s$, that can be retained at each step of the calculation.
As usual for NRG however, results converge quickly as function of $N_s$ for fixed $\Lambda$. For selected parameter values we tested $\Lambda=1.75, \ldots, 9$ and confirmed that the qualitative behavior of physical observables is independent of $\Lambda$, provided that $N_s$ is chosen sufficiently large.
We found that calculations with $N_s=2000$ for discretization parameter $\Lambda=2$--$3$ were fully converged down to temperature or energy scales $\sim 10^{-9}D$. Most of the calculations presented in the paper therefore  employed $\Lambda=3$ and $N_s=2000$.
Thermodynamic quantities are obtained in the standard fashion\cite{nrg_rev} from the NRG eigenstates and energies at each iteration.
However, since \textit{absolute} energies in NRG depend on the discretization parameter $\Lambda$, we also performed the extrapolation $\Lambda\to 1$ to determine the phase diagram, as discussed further below.


\section{Additional NRG results and analysis}

In this section we display NRG results which complement those of the main paper, and we provide some details for the analytical arguments.

\begin{figure}[t]
\begin{center}
\includegraphics[width=8.7cm]{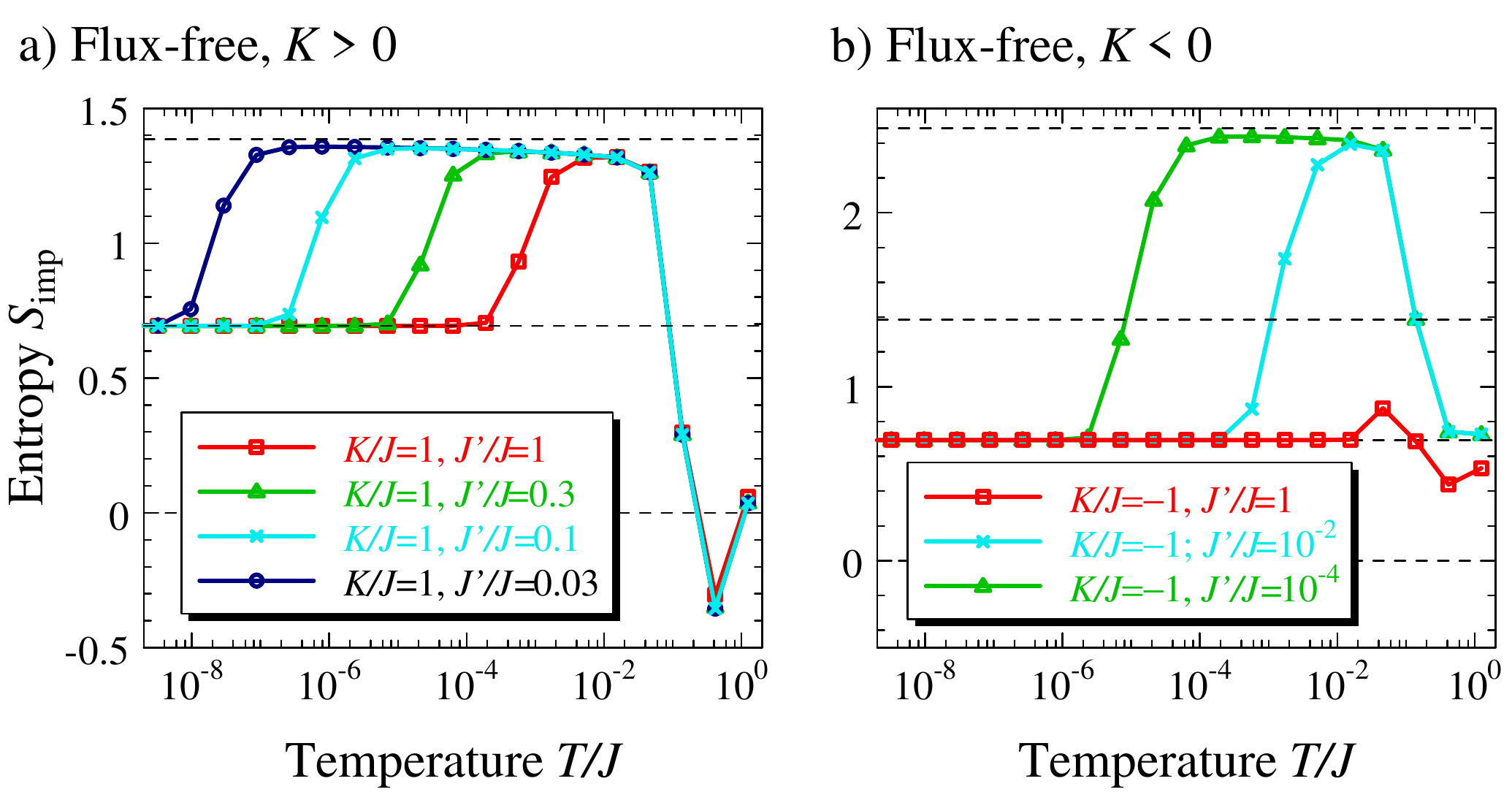}
\caption{\label{fig:s_noflux2}
NRG results for the impurity contribution to the total entropy, $\Simp(T)/k_B$ vs $T/J$ in the no-flux case. In contrast to Fig.~{\figsnoflux} of the main paper, we show here data for $J'\neq J$. The horizontal dashed lines indicate $\Simp=0$, $\ln 2$, $\ln 4 \approx 1.39$, and $\ln 12 \approx 2.48$.
}
\end{center}
\end{figure}

\begin{figure}[t]
\begin{center}
\includegraphics[width=8.7cm]{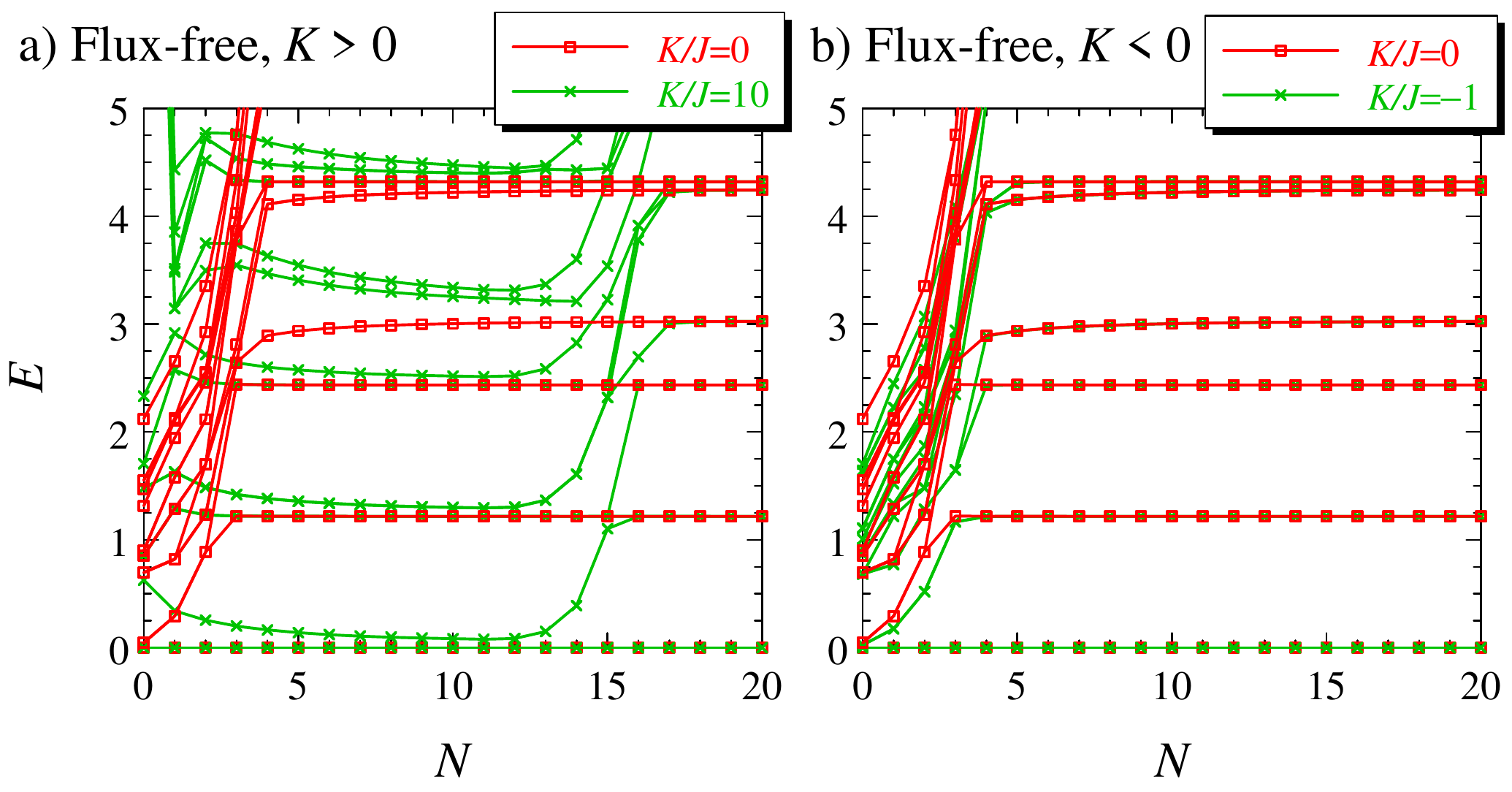}
\caption{\label{fig:flow_noflux}
Flow of the lowest NRG levels for the flux-free case and different values of $K$, illustrating that the LM fixed point corresponding to $K=0$ is reached irrespective of the initial $K$.
}
\end{center}
\end{figure}

\subsection{Flux-free case}

As before, we start with the flux-free sector of the Hilbert space. The impurity entropy $\Simp$ was shown in Fig.~{\figsnoflux}; Fig.~\ref{fig:s_noflux2} displays similar data which now include parameter sets with modified bulk couplings near the impurity, $J'\neq J$. Those parameters are useful for a comprehensive understanding of the qualitative RG flow as shown in Fig.~{\figrgflow} of the main paper.
In addition to $\Simp$, we show in Fig.~\ref{fig:flow_noflux} the flow of the lowest energy levels of the NRG Hamiltonian -- the NRG level pattern serves as a fixed-point fingerprint;\cite{nrg,nrg_rev} it is related to the finite-size spectrum. We note that, in the absence of a field, all levels are at least doubly degenerate because the three components of a composite pseudospin-$1/2$ operator constructed from $\hat{S}$ and the plaquette fluxes are conserved.\cite{dhochak15}

Figs.~\ref{fig:s_noflux2} and \ref{fig:flow_noflux} underline that all parameter sets with $|K|<\infty$ cause a flow to the LM fixed point with $\Simp^{\rm LM}=\ln 2$. Hence, there is no Kondo screening at any $K$ in the flux-free sector, at variance with the conclusions in Refs.~\onlinecite{dhochak10,dhochak15}. Both SVac and TVac appear as infrared unstable fixed points, with entropies of $\ln 4$ and $\ln 12$, respectively. Notably, these entropies are approached logarithmically slowly: The reason is that the bath DOS in the $s$-wave channel diverges with a logarithmic correction, $\rho(\w) \propto 1/(\w\ln^2\w)$, hence the fixed point is approached in a fashion similar to a marginally irrelevant perturbation.

\begin{figure}[t]
\begin{center}
\includegraphics[width=8.7cm]{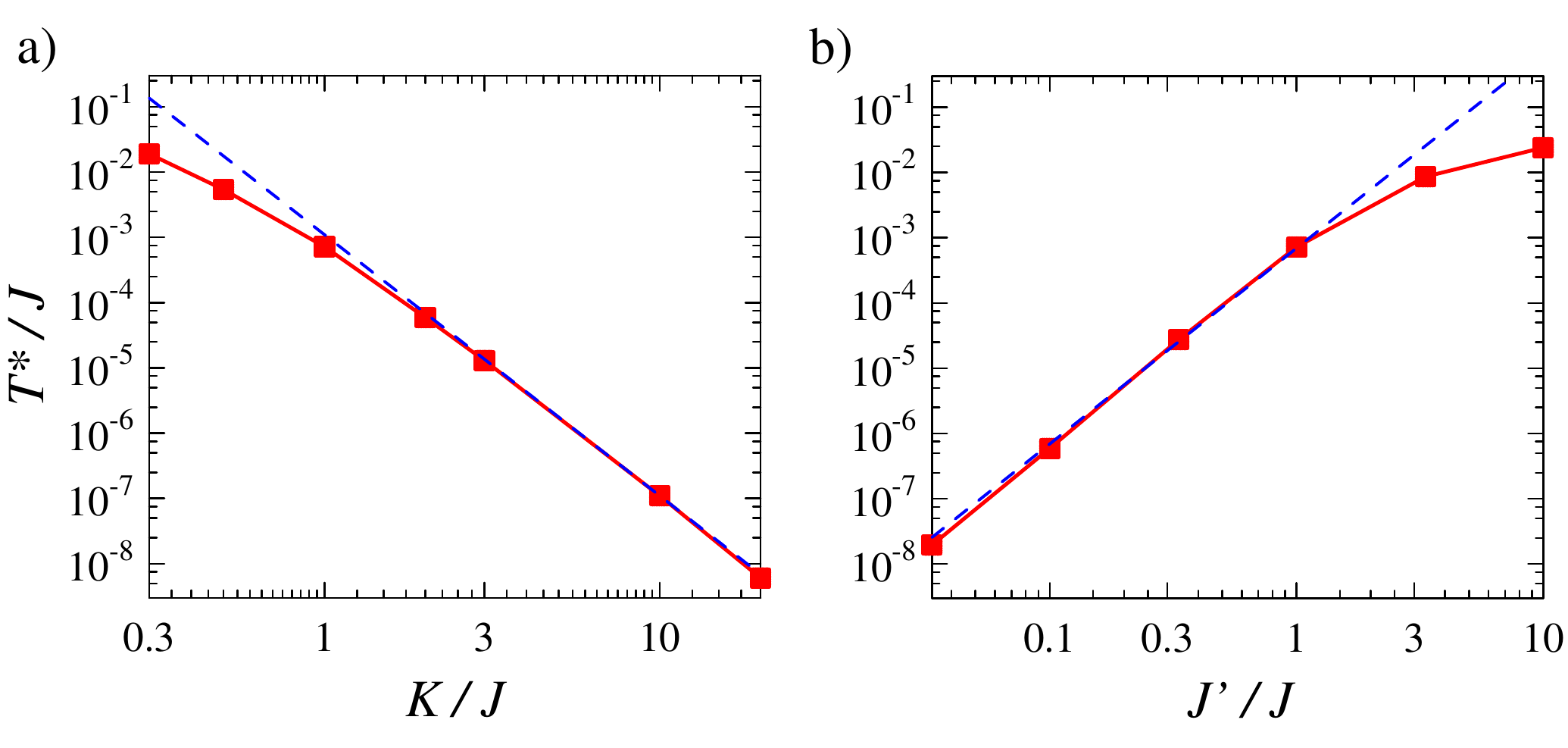}
\caption{\label{fig:tast_noflux}
Crossover scale $T^\ast$ for the flux-free case and positive $K$ as extracted from NRG data as shown in Figs.~{\figsnoflux}(a) and \ref{fig:s_noflux2}(a),
(a) as function of $K$ for $J'/J=1$ and
(b) as function of $J'$ for $K/J=1$.
$T^\ast$ has been defined via $\Simp(T^\ast)=1.0$.
The dashed lines indicate power-law fits:
(a) $T^\ast \propto 1/K^4$, (b) $T^\ast \propto J'^3$.
}
\end{center}
\end{figure}

The energy scale $T^\ast$ for the crossover from SVac to LM for the positive-$K$ case has a non-trivial parameter dependence. We have collected $T^\ast$ values in Fig.~\ref{fig:tast_noflux}; the fits demonstrate an approximate dependence $T^\ast \propto J'^3 J^2 / K^4$. Deviations occur for large $T^\ast$ where the asymptotic regime has not yet been reached; in addition small deviations are visible at small $T^\ast$ which we attribute to logarithmic corrections which arise from the logarithmic flow towards SVac.

The qualitative parameter dependence of $T^\ast$ can be deduced by analyzing the vicinity of the SVac fixed point. Although SVac consists of a non-degenerate impurity coupled to a bath (where site $0$ has been cut out), this situation is unstable because the bath DOS diverges, Eq.~\eqref{rho0}. This situation is similar to that of the fermionic pseudogap Kondo model at particle--hole (p-h) symmetry where the strong-coupling singlet fixed point is unstable.\cite{GBI,FV04} In both cases, a bath DOS for the Kondo impurity $\propto |\w|^r$ implies a bath DOS for the singlet $\propto |\w|^{-r}$ when site $0$ is removed. Power counting in the relevant Anderson model shows that the leading perturbation to the singlet fixed point has scaling dimension $(2r-1)$ for $1/2<r\leq1$. In the pseudogap Kondo model, this perturbation is essentially given by the inverse of the Kondo coupling $J_K$.
In our case, with $r=1$, the nature of the coupling implies that a third-order process is required, such that the (dimensionless) perturbation is $J'^xJ'^yJ'^z/(JK^2)$, with scaling dimension unity up to logarithmic corrections (see Ref.~\onlinecite{dhochak10} for related considerations).
Now, the singlet fixed point is destabilized once this perturbation is strong enough to break the singlet which happens on the scale $K^2/J^2$, resulting in $T^\ast/J \propto J'^3 J/K^4$ to logarithmic accuracy. This is analogous to the case of the pseudogap Kondo model at $r=1$ where the strong-coupling fixed point is destabilized at a scale\cite{GBI,AKMpc} $T^\ast\propto W^4/J_K^3$ where $W$ is the bandwidth of the fermionic bath.

\subsection{Impurity-flux case}

\begin{figure}[t]
\begin{center}
\includegraphics[width=8.7cm]{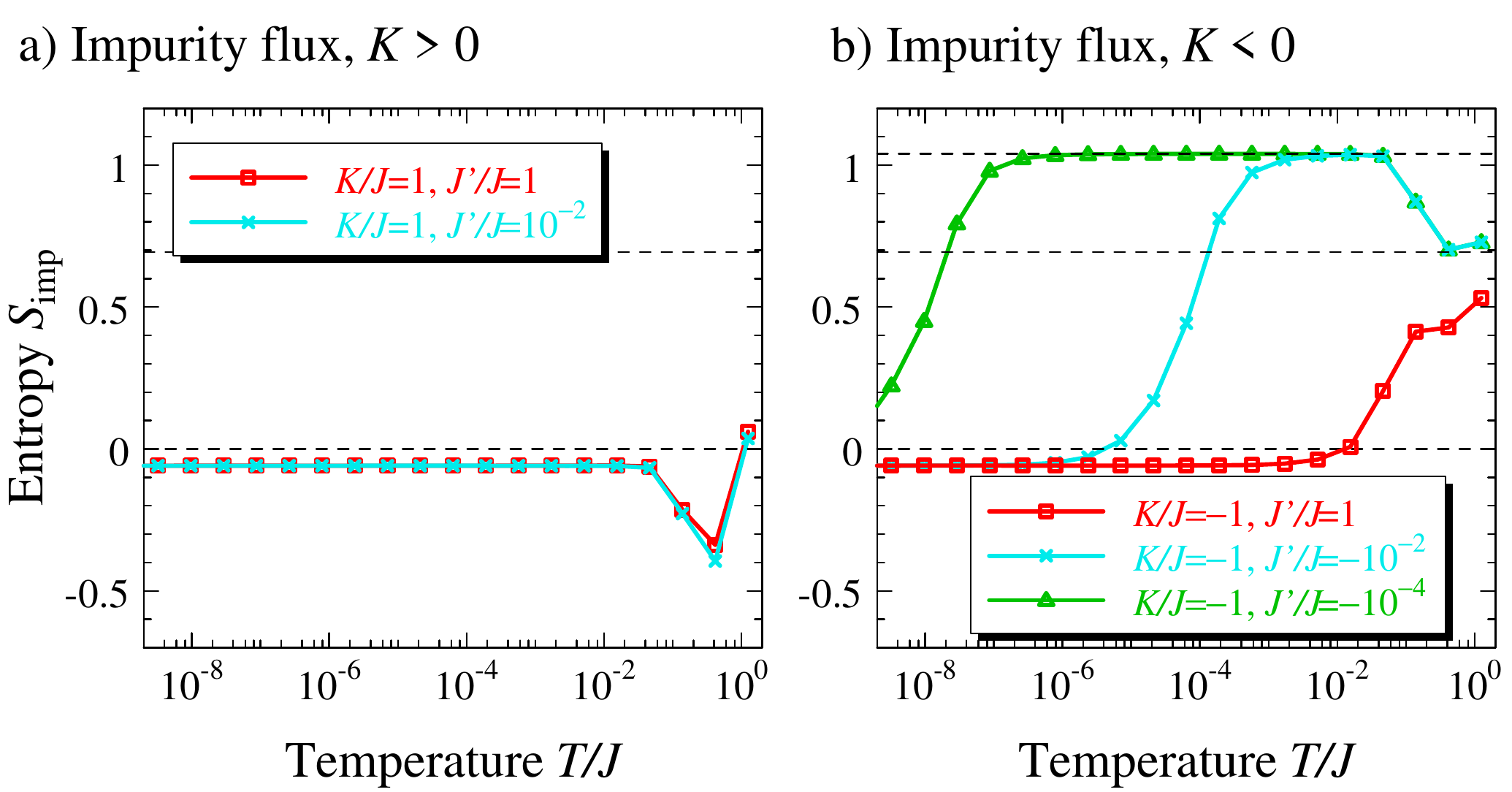}
\caption{\label{fig:s_flux2}
Impurity entropy as in Fig.~\ref{fig:s_noflux2}, but for the impurity-flux case and $J'\neq J$. The horizontal dashed lines indicate $\Simp=0$, $\ln 2$, and $\tfrac{3}{2} \ln 2$.
}
\end{center}
\end{figure}

\begin{figure}[t]
\begin{center}
\includegraphics[width=8.7cm]{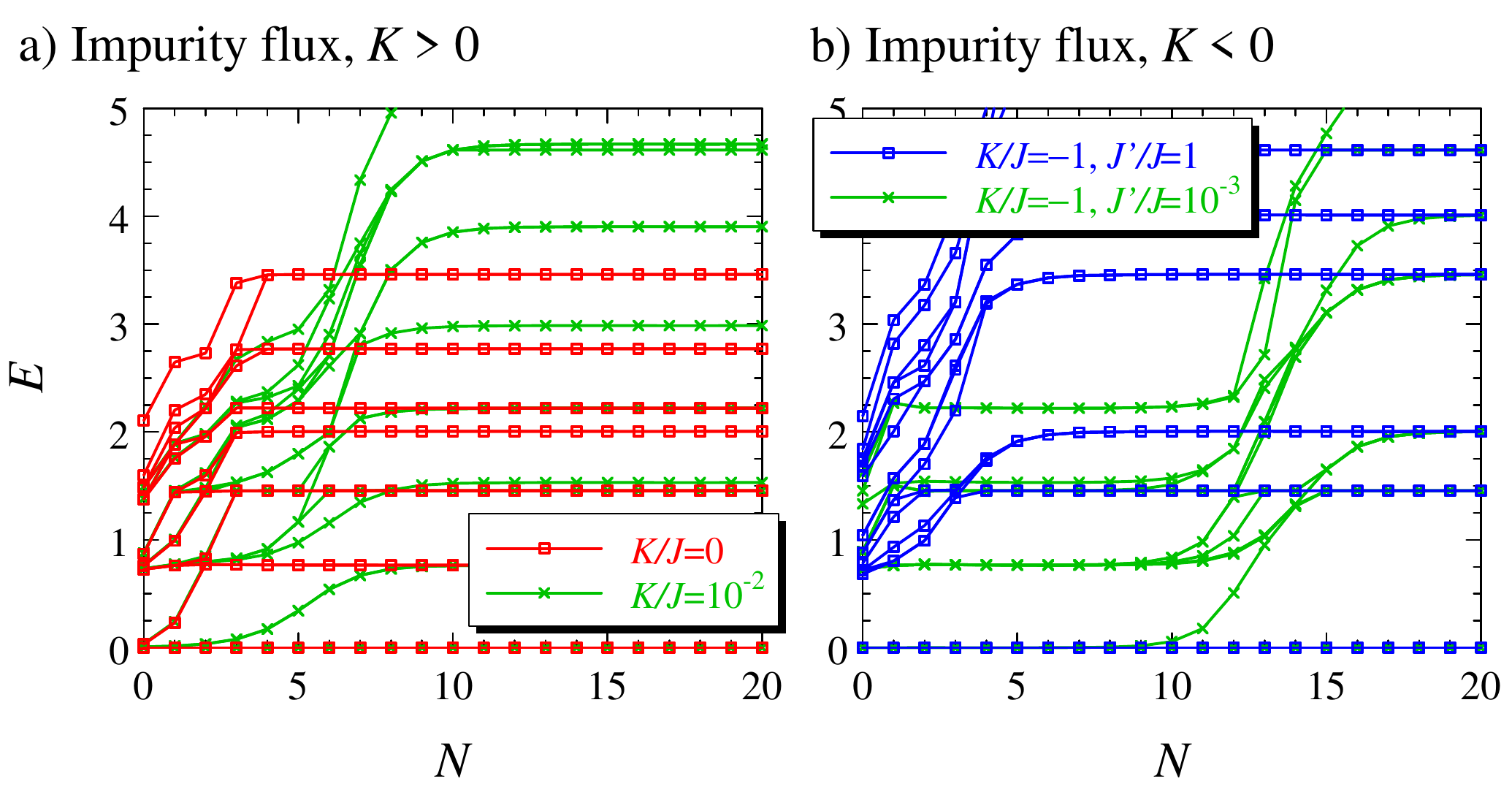}
\caption{\label{fig:flow_flux}
NRG level flow for the impurity-flux case and different parameters.
(a) The LM fixed point ($K=0$) is unstable for finite $K>0$, the system instead flows to SVac.
(b) For negative $K$ the system ultimately flows to TVac$'$, and for small $J'$ an intermediate regime corresponding to TVac is realized.
Note that level degeneracies of TVac are lifted upon flowing to TVac$'$;
further the level patterns of TVac$'$ and SVac are different (while $\Simp$ for both fixed points is identical).
}
\end{center}
\end{figure}

In the Hilbert-space sector with impurity flux, $W_I=-1$, the flow is directed towards SVac for $K>0$ and towards TVac$'$ for $K<0$. This is illustrated in Figs.~{\figsflux}, \ref{fig:s_flux2}, and \ref{fig:flow_flux}. While both SVac and TVac$'$ fixed points share the same impurity entropy, $\Simp = \tfrac{3}{2}\ln 2 - \ln 3$, their level pattern is clearly different, Fig.~\ref{fig:flow_flux}.

The crossover from TVac to TVac$'$ is documented in Figs.~\ref{fig:s_flux2}(b) and \ref{fig:flow_flux}(b). TVac displays an entropy of $\Simp=\tfrac{3}{2}\ln 2$ which is larger by $\ln 3$ than that of TVac$'$ -- this difference corresponds to the isolated spin-$1$ degree of freedom formed for $K=-\infty$, $J'=0$, consistent with the level degeneracies at TVac. Upon coupling this spin-$1$ to the bath via a finite $J'$, the degeneracies corresponding to spin-$1$ are lifted due to the lack of $\SUtwo$ spin symmetry in the Kitaev host. This level splitting occurs as a second-order perturbation, hence the crossover scale in Fig.~\ref{fig:s_flux2}(b)
is $T^\ast \propto J'^2/J$. 

\begin{figure}[t]
\begin{center}
\includegraphics[width=8.7cm]{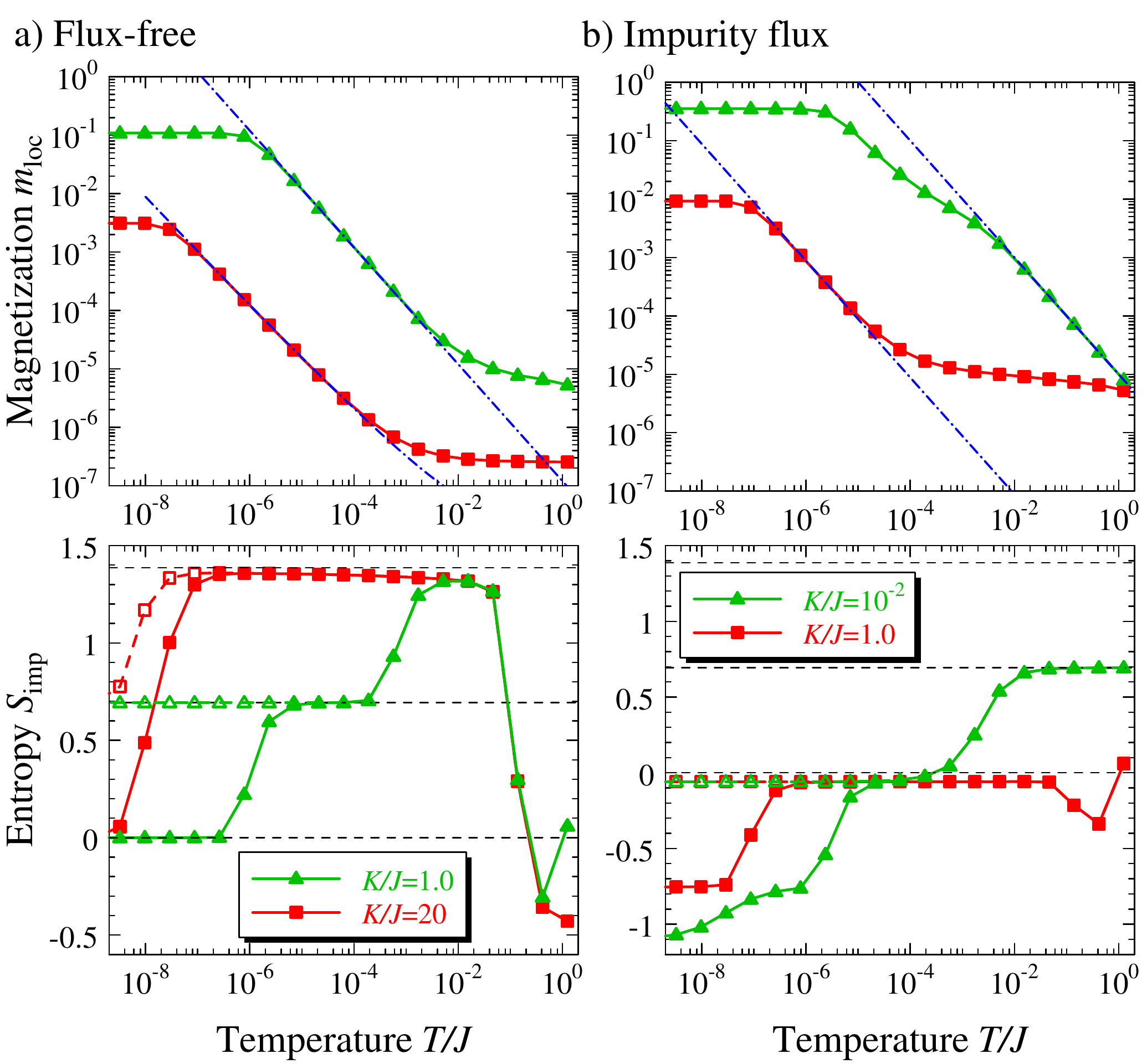}
\caption{\label{fig:m}
NRG results for impurity magnetization \cite{sunityfoot} and entropy for a field $\vec{h}\parallel z$.
Solid lines/full symbols show data for $h^z/J = 10^{-5}$; dashed lines/open symbols represent $h=0$ data for comparison. The dash-dot lines show specific fits to $\mloc(T)$.
(a) No-flux case. For $K/J=1$ the fit reflects the Curie law of the LM phase, while for $K/J=20$ the fit is of the form $1/(T\ln T)$ characteristic of SVac.
(b) Impurity-flux case. For $K/J=10^{-2}$ the fit describes the high-$T$ Curie law of LM, while for $K/J=1$ the fit is for the low-$T$ Curie law of the SVac phase.
}
\end{center}
\end{figure}

\begin{figure}[bt]
\begin{center}
\includegraphics[width=8.7cm]{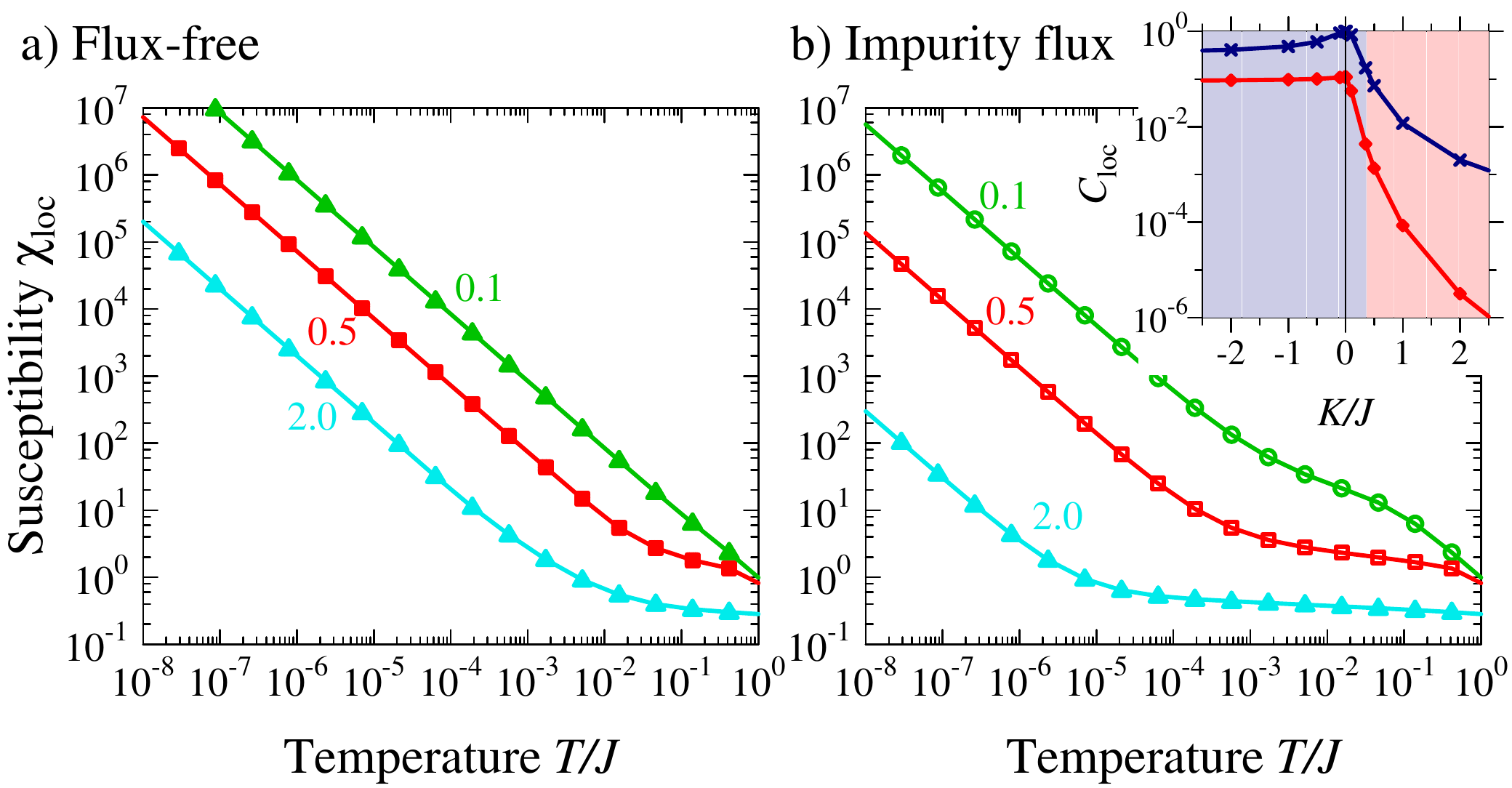}
\caption{\label{fig:chi}
NRG results for local susceptibility $\cloc(T)$ for $\vec{h}\parallel z$, obtained as $\cloc = \mloc/h$ at $h^z/J=10^{-9}$, for $J'=J$ and different $K/J$ as labelled.
(a) No-flux case.
(b) Impurity-flux case.
The inset shows the low-$T$ Curie constant as function of $K/J$ for the flux-free (impurity flux) sector in blue (red); the shading indicates which flux sector is the ground-state sector. At the phase boundary, the Curie constant drops by a factor $40$.
}
\end{center}
\end{figure}

\subsection{Magnetic field}

A local magnetic field $h$, applied to the Kondo spin only, can be easily integrated into the NRG algorithm. In contrast, a global magnetic field which also acts on the bulk Kitaev system spoils the model's solubility because the $\Ztwo$ fluxes are no longer conserved. Hence, we restrict ourselves to analyzing the effect of a local field.

Sample results for both the local magnetization,\cite{sunityfoot} $\mloc = \langle S^z \rangle$, and the impurity entropy $\Simp$ for a field applied along $z$ are displayed in Fig.~\ref{fig:m}. In the small-field limit, the magnetization allows to deduce the local susceptibility $\cloc=\mloc/h$, with results shown in Fig.~\ref{fig:chi}.

We start the discussion with the flux-free case. We know that the field-free system flows to the LM fixed point; consistent with this we observe a Curie law, $\cloc(T) = \Cloc/T$, at low temperature for any $K$, corresponding to an unscreened spin. The interaction with the bath causes a reduction of the local Curie constant $\Cloc$ from its free-spin value\cite{sunityfoot} $4S(S+1)/3=1$. For large values of $K$ the crossover temperature towards LM is small as detailed above, and an intermediate-temperature SVac regime appears. Its response is of the form\cite{willans10} $\chi \propto 1/(T\ln T)$; this quasi-free moment arises from the dangling gauge Majorana fermions $b_1^x$, $b_2^y$, $b_3^z$ together with the divergent bath DOS. Importantly, this response is located on the sites $1,2,3$ (strictly speaking, for $\vec{h}\parallel z$ the response arises from $b_3^z$ and is located at site $3$), and the Kondo-spin response $\cloc$ \textit{at} the SVac fixed point (i.e., for $K=\infty$) vanishes. However, for $K<\infty$ virtual excitations of the singlet (formed by the Kondo spin and the Kitaev spin at site 0) transmit the $1/(T\ln T)$ response to the Kondo spin. The impurity entropy is fully quenched at low $T$ once the magnetization reaches its low-$T$ saturation value.

In the impurity-flux case, the flow for positive $K$ is towards SVac. Its response is highly non-trivial: First, the $K=\infty$ case has a logarithmically divergent response $\chi\propto 1/\ln T$ on sites $1,2,3$ which arises from the dangling gauge Majoranas as above, but now together with a regular bath DOS.\cite{willans11}
Second, for $K<\infty$ a pair of the dangling gauge Majoranas induces an additional Curie response because they get coupled via virtual excitations of the singlet. Formally, we may use bond operators\cite{bondop} $s$, $t_\alpha$ to describe the four states of the dimer formed by the Kondo spin and the Kitaev spin at site $0$. For $J'=0$ its singlet ground state is $|s\rangle=s^\dagger|vac\rangle$.
The $J'$ pieces of the Hamiltonian take the following form:
\begin{equation}
J'^x \hat{\sigma}_0^x \hat{\sigma}_1^x = J'^x \hat{b}_1^x \hat{c}_1 (t_x^\dagger s + s^\dagger t_x + \ii t_y^\dagger t_z - \ii t_z^\dagger t_y)\,.
\end{equation}
In the spirit of perturbation theory around the $J'=0$ limit we can consider pieces of the wavefunction with virtual triplet excitations, obtained by repeated application of the $J'$ term of the Hamiltonian:
\begin{equation}
\label{wf1}
|\psi_0\rangle = (1 + \alpha \hat{b}_1^x \hat{c}_1 t_x^\dagger s + \beta \hat{b}_1^x \hat{c}_1 \hat{b}_2^y \hat{c}_2 t_z^\dagger s + \ldots) |s\rangle \otimes |{\rm kit}\rangle
\end{equation}
where $|{\rm kit}\rangle$ denotes the ground state of the Kitaev host with a vacancy, and $\alpha$ and $\beta$ are non-zero coefficients which are suppressed with powers of $J'/K$.
Importantly, if we evaluate the local moment with this wavefunction we find:
\begin{align}
\langle\psi_0|\hat{S}^z|\psi_0\rangle &= \langle\psi_0|-t_z^\dagger s - s^\dagger t_z + \ii t_x^\dagger t_y - \ii t_y^\dagger t_x)|\psi_0\rangle \notag\\
&= \beta \langle {\rm kit}| \hat{c}_1\hat{c}_2|{\rm kit}\rangle \langle \hat{b}_1^x\hat{b}_2^y\rangle + \ldots
\end{align}
The expectation value $\langle {\rm kit}| \hat{c}_1\hat{c}_2|{\rm kit}\rangle$ is finite for the hopping problem at hand, hence we have the remarkable result
\begin{equation}
\langle\hat{S}^z\rangle \propto \langle \hat{b}_1^x\hat{b}_2^y\rangle
\end{equation}
This implies that the (originally uncoupled) gauge Majorana fermions $\hat{b}_1^x$ and $\hat{b}_2^y$ form a spontaneous moment in $z$ direction. This couples to $h^z$ and causes a Curie response. As a result, the local response for $K<\infty$ is a superposition of a $1/\ln T$ piece and a Curie piece. For $\vec{h}\parallel z$ the former is induced by $b_3^z$ and the latter by $b_1^x$ and $b_2^y$. Due to the small $\Cloc$ of the Curie piece this will dominate only at low $T$, and an intermediate regime of logarithmic response remains visible, Fig.~\ref{fig:chi}(b).

Finally, if $K$ is very small, an intermediate LM regime is visible which displays a Curie response with $\Cloc=1$, Fig.~\ref{fig:m}(b).

The impurity entropy in the impurity-flux case shows a field-induced quench of the $\tfrac{3}{2} \ln 2$ contribution from the dangling gauge Majoranas in the low-$T$ limit, such that $\Simp$ becomes $(-\ln 3)$, Fig.~\ref{fig:m}(b) -- recall that $(-\ln 3)$ arises from removing the flux degeneracy of the uncoupled bath. However, depending on the field direction, the individual Majorana contributions are quenched at different temperatures. This is natural because -- as just explained -- the Majoranas play different roles in the moment formation. More precisely, for $\vec{h}\parallel z$ the entropy of $b_1^x$ and $b_2^y$ is quenched below the temperature where the Curie moment reaches ``saturation'', whereas quenching the remaining $\tfrac{1}{2}\ln 2$ from $b_3^z$ requires the coupling to the NRG bath and happens at much lower $T$, Fig.~\ref{fig:m}(b).

For negative $K$ the flow in the impurity-flux case is towards TVac$'$. Its field response (not shown) is similar to that of SVac, with a superposition of logarithmic and Curie contributions, the latter now with $\Cloc$ of order unity because the equivalent of the $\alpha$ and $\beta$ coefficients in Eq.~\eqref{wf1} are not suppressed with $J'/|K|$. We recall that TVac$'$ does not display a free spin-$1$ moment (as opposed to TVac) because the coupling to the Kitaev host lifts the spin degeneracy.

In summary, all stable phases of the Kitaev Kondo model display a Curie response, but with drastically different Curie constants $\Cloc$. This is summarized in the inset of Fig.~\ref{fig:chi}: While $\Cloc$ tends to decrease with increasing $|K|$ in both flux sectors, for $K>0$ it is smaller by one or more orders of magnitude in the impurity-flux case as compared to the flux-free one. Hence, the first-order quantum phase transition between the two flux sectors is accompanied by a change in $\Cloc$ by about a factor $40$, which should be detectable in a suitable NMR experiment.

\subsection{Flux transition}

To detect the first-order quantum phase transition between the two flux sectors, an accurate comparison of their ground-state energies is required. The energy difference, $\Delta E = E_{\rm flux} - E_{\rm no flux}$, has two contributions, one arising from the NRG bath (i.e. from the problem with $J'=K=0$) and one arising from the quantum impurity problem itself -- the latter is obtained from the NRG algorithm. The bath contribution $\Delta E_{\text{bath}}$ is equivalent to the flux-binding energy of a vacancy and has been determined in Ref.~\onlinecite{willans11}:
\begin{equation}
(E_{\rm flux} - E_{\rm no flux})_{\text{bath}} = -0.027 J\,.
\end{equation}

\begin{figure}[t]
\begin{center}
\includegraphics[width=8.7cm]{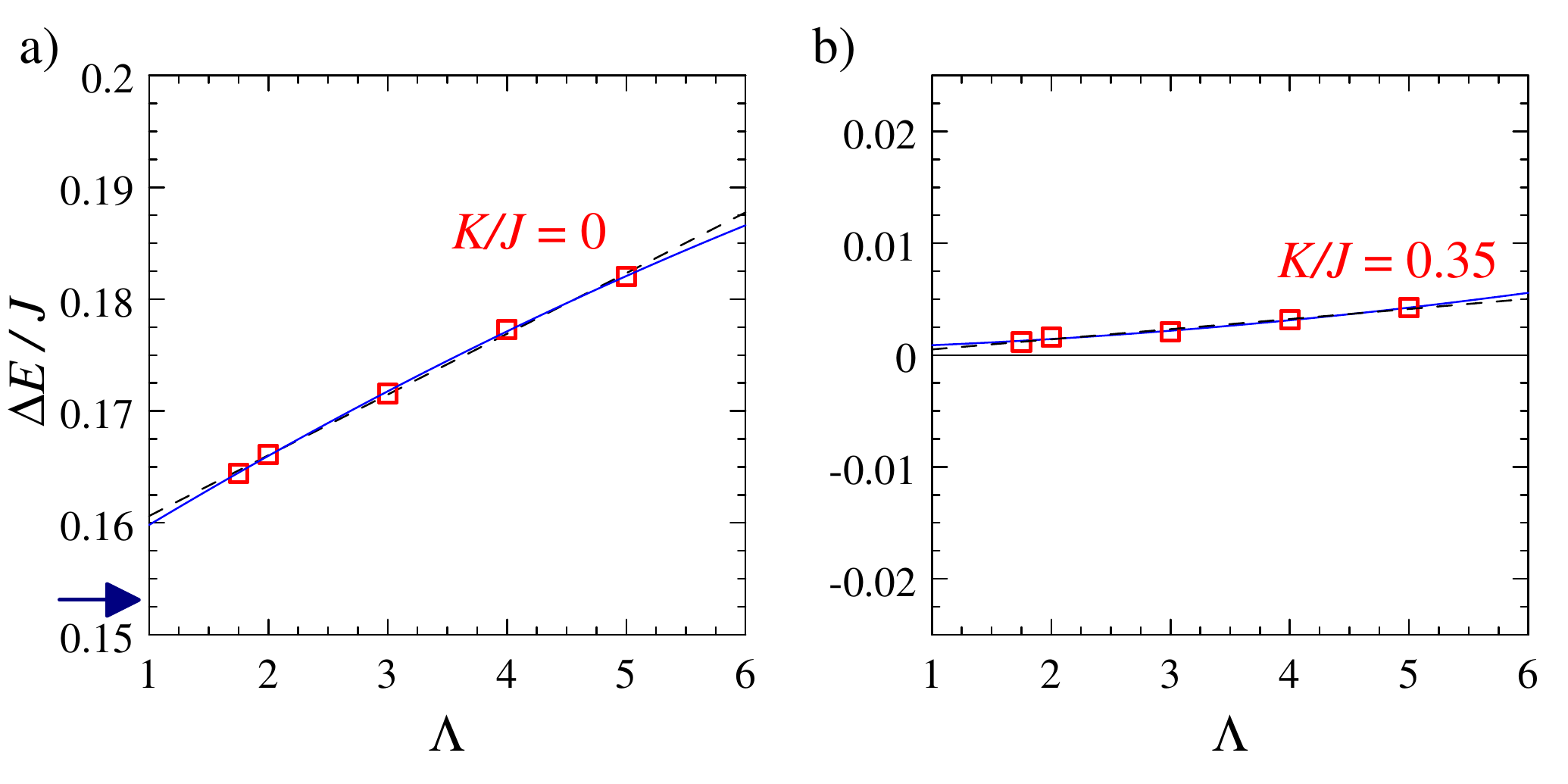}
\caption{\label{fig:ediff_lam}
$\Lambda$ exptrapolation for the energy difference between impurity-flux and no-flux ground states, $\Delta E =  E_{\text{flux}}-E_{\text{noflux}}$. The bath contribution\cite{willans11} $\Delta E_{\text{bath}}=-0.027J$ is taken into account.
(a) $K=0$: The known result\cite{willans11} $\Delta E=0.153J$ is marked by an arrow.
(b) $K/J=0.35$, close to the first-order QPT.
The dashed (solid) lines indicate a linear (quadratic) fit.
}
\end{center}
\end{figure}

Obtaining an estimate of the NRG piece $\Delta E_{\text{NRG}}$ requires to calculate the ground-state energies for fixed model parameters and different values of the NRG discretization parameter $\Lambda$ and then to extrapolate $\Delta E_{\text{NRG}}$ to the (formally exact) limit of $\Lambda=1$. Such an extrapolation is shown in Fig.~\ref{fig:ediff_lam}, with the $\Lambda$ dependence being approximately linear.

To gauge the accuracy, we can make use of the fact that the model at $J'=J$, $K=0$ represents the bulk Kitaev model: Here, the energy cost of a single plaquette flux in the thermodynamic limit has been calculated to be\cite{willans11} $\Delta E=0.153J$. The $\Lambda$ extrapolation yields a value of $0.160J$, i.e., a deviation of less than one percent of $J$ -- the good agreement can be considered as consistency check for our NRG procedure. The remaining deviation is rooted in the NRG algorithm, as the logarithmic discretization of the bath spectrum is inherently imprecise at elevated energies which influences spectra-integrated quantitities such as the total energy.\cite{nrg_rev}
We note that $\Delta E$ in the limit $K\to\infty$ is accurate by construction: Here $\Delta E_{\text{NRG}}=0$ because the singlet is cut out from the system, such that $\Delta E = \Delta E_{\text{bath}}=-0.027J$.
In order to make efficient use of the known limits, we have generated the data in Fig.~{\figen} of the main paper by rescaling $\Delta E_{\text{NRG}}$ by a constant factor to obtain the correct $\Delta E$ at $K=0$.


\section{Comparison to earlier work}

Given that our results differ in a number of important aspects from those obtained in Refs.~\onlinecite{dhochak10,dhochak15} on the same model, we highlight and analyze the differences in the following.

Refs.~\onlinecite{dhochak10,dhochak15} concluded that the flux-free sector displays a quantum phase transition between unscreened and screened phases; we have shown that such a transition is absent. The conclusion of Refs.~\onlinecite{dhochak10,dhochak15} is based, on the one hand, on a weak-coupling RG. However, the critical fixed point predicted by the RG is outside the weak-coupling regime and does not exist. This has been well studied for the pseudogap Kondo model with bath exponents $r>1/2$ where weak-coupling RG incorrectly predicts a quantum phase transition at p-h symmetry as well.\cite{withoff,GBI,FV04} On the other hand, Ref.~\onlinecite{dhochak15} argued the strong-coupling fixed point (SVac in our notation) to be generically stable. However, in the flux-free case the singular DOS of the environment surrounding the singlet, ignored in Ref.~\onlinecite{dhochak15}, destabilizes the SVac fixed point, again similar to what happens in the p-h-symmetric pseudogap Kondo model.\cite{GBI,FV04}

Ref.~\onlinecite{dhochak15} argued that the SVac fixed point has a residual entropy of $\tfrac{1}{2} \ln 2$ from a single Majorana zero mode, akin to the two-channel Kondo effect. Our NRG results instead show that $\Simp$ is more complicated: It has contributions from \textit{three} Majorana zero modes ($b_1^x$, $b_2^y$, $b_3^z$); further the flux-free sector has an additional $\tfrac{1}{2} \ln 2$ from the singular bath DOS, while the impurity-flux sector has a $(-\ln 3)$ contribution due to the quenched flux degeneracy. We believe the fact that, e.g., $\ii b_1^x b_2^y$ is conserved at $K=\infty$ cannot be used to discard its entropy contribution, as done in Ref.~\onlinecite{dhochak15}. We have also shown that, in the impurity-flux case, the SVac phase has a residual Curie term in the susceptibility for $K<\infty$; this has been missed in Ref.~\onlinecite{dhochak15} and is also different from the two-channel Kondo phenomenology.

Finally, numerical results in Ref.~\onlinecite{dhochak15} placed the flux-binding transition for antiferromagnetic coupling\cite{dhochakpc} at $K_c\approx0.1 J$ and further suggested that there is a flux-binding transition also in the case of ferromagnetic Kondo coupling (and zero applied field). However, these results were obtained for extremely small host systems and likely suffered from strong finite-size effects. Our results obtained in the thermodynamic limit indicate a single transition at $K_c\approx 0.35J$ instead.
